\documentclass{aa} 
\usepackage{graphicx}
\usepackage{color}
\usepackage{booktabs}
\usepackage{makecell}
\usepackage{longtable}
\usepackage{ulem}
\usepackage{lscape}
\usepackage{amsmath,amssymb}
\usepackage{array,multirow,xcolor}
\usepackage{xurl}
\usepackage{placeins}
\usepackage{stfloats}

\newcommand{\kms}{\,\hbox{\hbox{km}\,\hbox{s}$^{-1}$}}
\newcommand{\mins}{\,\hbox{\hbox{min}}}

\newcommand{\htwo}{\,\hbox{$\rm{H_ 2}$}}

\newcommand{\cooz}{\hbox{$\rm CO(1$-$0)$}}
\newcommand{\coto}{\hbox{$\rm CO(2$-$1)$}}
\newcommand{\cott}{\hbox{$\rm CO(3$-$2)$}}
\newcommand{\coft}{\hbox{$\rm CO(4$-$3)$}}

\usepackage[varg]{txfonts}

\usepackage{hyperref}

\begin{document} 

   \title{CO in the ALMA Radio-source Catalogue (ARC): The molecular gas content of radio galaxies as a function of redshift \thanks{Reduced datacubes are only available in electronic form
at the CDS via anonymous ftp to cdsarc.u-strasbg.fr (130.79.128.5)
or via http://cdsweb.u-strasbg.fr/cgi-bin/qcat?J/A+A/
}}
   \titlerunning{Molecular gas content of radio galaxies}
   \authorrunning{Audibert et al. }
   
   \author{ A. Audibert\inst{1,2,3}, K.\,M. Dasyra\,\inst{4,1}, M. Papachristou\,\inst{4,1}, J.\,A. Fern\'andez-Ontiveros\,\inst{1,5}, I. Ruffa\inst{1,6}, 
   L.~Bisigello\inst{7,8}, 
   F.~Combes\inst{9,10},
   P.~Salomé\inst{9},
   C.~Gruppioni\inst{7}}

   \institute{
   Institute of Astronomy, Astrophysics, Space Applications, \& Remote Sensing, National Observatory of Athens, Palaia Penteli, 15236, Athens, Greece
   \and
   Instituto de Astrof\' isica de Canarias, Calle V\' ia L\'actea, s/n, E-38205, La Laguna, Tenerife, Spain\\
   \email{anelise.audibert@iac.es}
   \and
   Departamento de Astrof\'isica, Universidad de La Laguna, E-38206, La Laguna, Tenerife, Spain
   \and
   Department of Astrophysics, Astronomy \& Mechanics, Faculty of Physics, University of Athens, Panepistimiopolis Zografos 15784, Greece
   \and
   Centro de Estudios de F\'isica del Cosmos de Arag\'on (CEFCA), Plaza San Juan 1, E--44001 Teruel, Spain
   \and
   Cardiff Hub for Astrophysics Research \& Technology, School of Physics \& Astronomy, Cardiff University, Queens Buildings, The Parade, Cardiff CF24 3AA, UK
   \and 
   Istituto Nazionale di Astrofisica: Osservatorio di Astrofisica e Scienza dello Spazio di Bologna, Via Gobetti 93/3, 40129, Bologna, Italy 
   \and
    Dipartimento di Fisica e Astronomia, Università di Padova, Vicolo dell’Osservatorio, 3, I-35122, Padova, Italy
   \and
   Observatoire de Paris, LERMA, PSL University, Sorbonne University, CNRS, Paris, France 
   \and
   Coll{\`e}ge de France, 11 Pl. Marcelin Berthelot, 75231, Paris 
   }

   \date{Received 29 March 2022 / Accepted 12 August 2022}

  \abstract
  { To evaluate the role of radio activity in galaxy evolution, we designed a large archival CO survey of radio galaxies (RGs) to determine their molecular gas masses at different epochs. We used a sample of 120 RGs representative of the NVSS 1.4\,GHz survey, when flux limited at 0.4\,Jy. Of those, 66 galaxies belonged to the ALMA Radio-source Catalogue (ARC) of calibrators and had spectral window tunings around CO (1-0), (2-1), (3-2), or (4-3). We reduced their ALMA data, determined their \htwo\ mass contents, and combined the results with similar results for the remaining 54 galaxies from the literature. We found that, while at all epochs the majority of RGs have undetectable reservoirs, there is a rapid increase in the  \htwo\ mass content of the CO-detected RGs with $z$. At 1$<$$z$$<$2.5, one-fourth of the RGs have at least as much molecular gas as simulations would indicate for a typical halo mass of that epoch. These galaxies plausibly have ``normal'' or even starbursty hosts. Overall, reservoirs of $10^7 \lesssim M_{\rm H_2} \lesssim 10^{10} M_{\odot}$ are seen at $z<$0.3, and $10^{10} \lesssim M_{\rm H_2} \lesssim 10^{12} M_{\rm \odot}$ at $z>$1. Taking into account the completeness correction of the sample, we created the corresponding H$_2$ mass functions at 0.005$<$$z$$<$0.3 and 1$<$$z$$<$2.5.
  The local mass function
  reveals that the number density of low-z RGs with detectable molecular gas reservoirs is only a little lower (a factor of $\sim$4) than that of pure
  (or little star-forming) type 1 and 2 AGN in simulations. At 1$<$$z$$<$2.5, there is a significant decrease in the number density of high-z RGs due to the rarity of bright radio galaxies. An estimate for the missing faint RGs would, nonetheless, bring populations close again. Finally, we find that the volume density of molecular gas locked up in the brightest 1/5000-1/7000 RGs is similar in the examined $z$ bins. This result likely indicates that the inflow rate on one hand and the star-formation depletion rate plus the jet-driven expulsion rate on the other hand counteract each other in the most luminous RGs of each epoch.

  }



   \maketitle
%

\section{Introduction}
\label{sec:intro}

Feedback from active galactic nuclei (AGN) is widely believed to be a crucial element of the mass assembly of galaxies: with feedback acting on gas and stars forming from gas, feedback regulates the stellar mass growth of galaxies \citep{bower06,croton06,dimatteo08,sijacki07,dubois16, cielo18}.  By leading to the acceleration or heating of the gas in the interstellar medium (ISM), feedback is 
associated with winds and outflows that can even eject gas outside of galaxies.
 
It can also delay the deposition of new gas from the intracluster medium (ICM) or intergalactic medium (IGM) onto galaxies. 
Both effects help the quenching of star formation through the lack of fuel for accretion, simultaneously terminating the black hole activity \citep{alatalo15,lanz16}. 

It is widely believed that AGN feedback operates in a bimodal manner. The quasar mode, also called radiative or wind mode, occurs through wide-angle winds driven by radiative pressure in luminous AGN with accretion rates close to Eddington. This mode can be met at any redshift $z$, and it is very common in high-$z$, young QSOs \citep{king03,harrison17,bieri17}. The radio (or kinetic) mode involves the launch of collimated relativistic jet of particles (i.e.\,radio jet), and occurs mainly in low-luminosity AGN (<0.01L$_{\rm Edd}$). Radio jets launched by supermassive black holes (SMBH) are seen to have an impact even on large scales, e.g., of tens of kpc, delaying cooling flows from the ICM/IGM and the formation of extremely massive galaxies \citep{fabian12, namara12}. This is characteristically seen at the centers of fairly low $z$ clusters and groups. Of course, they can also have a strong impact on (sub-)kpc scales, altering the properties of the ISM and driving molecular outflows \citep{morganti15,dasyra16,fotopoulou19, oosterloo19,aalto20,juan20}, and even in low-luminosity AGN \citep{combes13,santi14,ane19,ruffa22}.  
The observational findings are supported by 3D hydrodynamical simulations of relativistic jets coupling in inhomogeneous and clumpy ISM \citep{wagner11,wagner12, muk18}, showing that propagating radio jets can gradually disperse gas clouds and create a cocoon of shocked material associated with multi-phase outflows. As this jet-driven bubble expands, it also leads to the dispersal of molecular clouds. The final feedback impact depends on the jet kinetic energy and the distribution of the clumpy medium defining the jet's path of minimum resistance.

Yet, a systematic and statistically meaningful observational measurement of the number of radio galaxies (RGs) with significant gas reservoirs, upon which the radio-mode feedback could be operating, is missing. So far, it has been shown that most of the local RGs have low gas content compared to spirals or infrared selected galaxies (e.g., Ruffa et al. 2022). High-$z$ CO surveys of RGs started in the 1990s \citep{evans96} and eventually led to the discovery  of large reservoirs of molecular gas (with $M_{\rm H_2}\sim10^{10}-10^{11}M_\odot$) in some powerful  RGs at $z\gtrsim$2 or gas-depleted, dead hosts in other \citep{scoville97,papa00,breuck03, breuck05,greve04, nesvadba08,emonts11,emonts14, castignani19}. The increased sensibility achieved with the advent of interferometric facilities such as the Atacama Large Millimeter Array  (ALMA) and the Northern Extended Millimeter Array (NOEMA) nowadays offer a great opportunity to study the evolution of the molecular gas in low and high-$z$ RGs.

In this paper, we present a new analysis of observations of CO molecular transitions from rotational numbers J=1-0 to J=4-3 performed with ALMA for RGs up to $z<$2.5. To build a statistically significant sample, we analyzed the CO emission in a sub-sample of RGs from the ALMA Calibrator Source Catalog, which were selected to be representative of the NRAO/VLA Sky Survey (NVSS) catalogue \citep{condon98}, with respect to its $z$ and  1.4 GHz flux distribution, down to a limit of 0.4 Jy. Complementing these with literature data, we carried out a large and complete archival CO survey of molecular gas in RGs. A main aim of our work was to analyze the evolution of the gas content of RGs with $z$, which is highly needed for an accurate benchmarking of the cosmological simulations. Such analysis also lead us to construct for the first time the CO luminosity function of RGs at low and high $z$. Another goal was to determine the occurrence of molecular outflows in the RGs in our sample.

The paper is structured as follows: in Section~\ref{sec:sample} we describe the statistics used to reproduce a representative sample of NVSS radio galaxies. In Section~\ref{sec:obs} we describe the data reduction of the ALMA observations. Main results from the analysis of the literature and ALMA observations are presented in Section~\ref{sec:results}. The presentation and discussion of the mass and luminosity functions are described Section~\ref{sec:lco}, followed by conclusions. In this paper we adopt a flat $\Lambda$CDM cosmological
model with H$_{\rm 0}$=70\kms Mpc$^{-1}$, $\Omega_{\rm M}$=0.3, and $\Omega_{\rm \Lambda}$=0.7.

\color{black}
\section{Radio galaxies with ALMA observations}
\label{sec:sample}
\color{black}
\subsection{ ALMA data mining }
\label{subsec:sample_arc}

The selection of galaxies for our survey started from the ALMA Radio Source Catalogue (ARC), which  comprises a list of mm and sub-mm bright objects that can serve as (flux, phase, bandpass, etc) calibrators. This catalogue includes a compilation of catalogues from several facilities other than ALMA, including the Very Large Array Calibrator Manual, Submillimeter Array and Atacama Compact Array catalogues, the Parkes survey, and the Combined Radio All-Sky Targeted Eight-GHz Survey. As such, the ensemble of the ARC covers the full sky, but in a non-homogeneous manner.

The ARC catalogue was downloaded from the European Southern Observatory (ESO) archival interface \footnote{\url{https://almascience.eso.org/sc/}}. Each catalogue entry corresponds to observations of a single object in a single band. More specifically, each entry shows the latest flux measurement of each object in a given band, with the exception of band 3. For band 3, two entries are provided, presenting the latest flux for each sideband. As of 01/08/2020, the downloaded catalogue contained 8679 entries. The number of unique objects (deduced by removing the information on the bands) was 3362. For some of the objects, the spectral windows of the calibrations covered the frequencies of molecular line transitions. Therefore, additional science on, e.g., the detection of CO lines, was feasible with the calibration data. Some of the sources in this catalogue were also observed as science targets. The combination of calibration and science related data led to a wealth of information that enabled our archival survey to be designed.

For each ARC source we used the "astroquery" tool \citep{astroquery} to find the available ALMA observation searching within a radius of 3\arcsec\ from the registered sky coordinates. Additionally, we iterated over all names given for each ARC catalogue entry in order to cross-check detection rates. In total, we found recorded 25827 observations from 1562 unique sources.

To determine whether a CO transition was in the spectral window of an observation, we needed the redshift of the observed source. For this purpose, we first queried the NASA Extragalactic Database (NED), Simbad, and Vizier using "astroquery". We performed an extra quality assessment by only keeping spectroscopically determined $z$ values, and in fact, by manually
examining whether each spectroscopic $z$ was measured using two or more spectral lines. Once having the $z$ information, we only kept observations of sources with CO(1-0), CO(2-1), CO(3-2), or CO(4-3) coverage in a spectral window. A total of 675 sources had at least one CO line with any integration time. We kept only sources with a minimum integration time of $5 \mins$ to ensure that the data analysis (e.g., the derivation of error bars) is meaningful.

As a next step, we had to ensure that only galaxies with radio emission associated with accretion onto black holes were kept. As the goal of our survey is to investigate the role of radio-mode feedback on galaxy evolution, all objects with radio emission associated with star formation activity had to be eradicated. For this reason, we imposed the 1.4GHz-to-24$\mu$m flux criterion of \citet{bonzini13}, identifying radio emission in excess of what supernovae can produce in star-forming galaxies via the quantity $q_{24}=\log(S_{\rm 24\mu m}/S_{1.4GHz})$, which has to be $<$0.5\footnote{The ``classical" definition of radio loudness via the $R$ indicator, the ratio between the rest-frame radio-to-optical flux density with typical values of R$\sim$10 characterizing RGs \citep{kellermann89}, is often insufficient to identify radio-quiet (RQ) objects, because both star-forming and RQ galaxies can have low R values.}.

The 1.4\,GHz fluxes of our sources were obtained directly from radio surveys, when available. Otherwise, we interpolated their values from
the nearest available radio data, fitting the radio data with a power law in $\nu F_\nu$.  In total, this requirement left us with 204 sources.  From here onward, we will refer to all radio galaxies with ALMA CO observations as the  CO-ARC catalogue.

To serve as calibrators, the CO-ARC sources are point sources at millimeter and radio wavelengths (such as quasars or blazars).  To also include RGs with spatially-resolved, extended radio jets, we performed an extensive bibliographic search, which provided us with a pool of either single-object or dedicated samples with CO observations \citep[e.g.][]{lim00,evans05,ocana10,ruffa19b,russell19, dabhade20}. The $q_{24}$ criterion described above was then applied to the selected literature sources, leaving us with a total of 152 galaxies which were then considered suitable to complement our CO-ARC sample. Hereafter, we will refer to the joint CO-ARC and bibliographic super-sample of 356 galaxies as the {\it extended} CO-ARC.

\begin{figure}[h!]
\resizebox{\hsize}{!}{\includegraphics{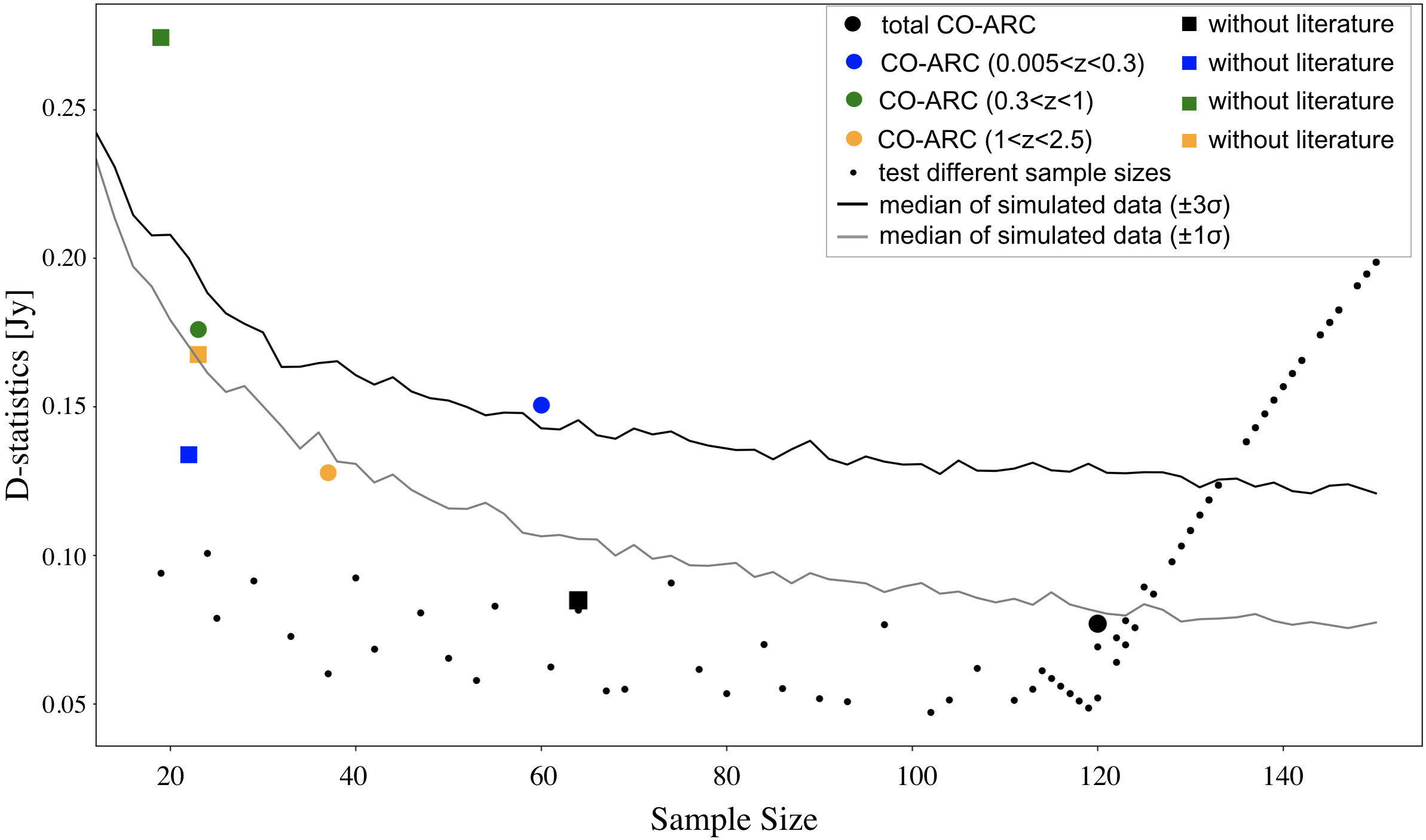}}\caption{D-statistic, i.e., maximum distance between two cumulative distributions used for the sample selection versus the sample size . Solid lines show a theoretical comparison of the NVSS and a shuffled NVSS samples with 10 and 30\%  flux variations. Small black dots show how the sample D-statistic changes as a function of the number of bootstrapped sources (see text). The large circles show the final numbers for the CO-ARC (black) and for individual $z$ ranges (color). The same analysis is shown for the CO-ARC only using the ALMA calibrators (without including the literature sources) is indicated as squares and equivalent color code for the redshift bin. We reach an optimal sample size of 120 (CO-ARC in black circle) for a D-statistics smaller than 10\% and converging with the values computed to the theoretical shuffled NVSS sample down to 10\% flux variation.}
\label{fig:convergence}
\end{figure}

\subsection{Selection of a radio-representative sample}
\label{subsec:sample_parent}

Our next task was to find a sub-sample of the {\it extended} CO-ARC sample that is representative of a radio galaxy survey. As parent radio catalogue, we used the NVSS 1.4\,GHz imaging survey, performed by the Very Large Array (VLA)  \citep{condon98}. The NVSS covers a large fraction of the sky for declinations above $\delta>-$40$^\circ$, and it is fairly deep in sensitivity,  recovering 1.4 GHz flux densities as low as $\sim$2\,mJy. This limit ensures completeness at the levels we are interested in examining, of a few hundred mJy, given the 1.4 GHz fluxes of the CO-ARC sources.  In fact,
we also examined a few other catalogues as potential parent surveys of ours:
the Fourth Cambridge Survey (4C) at 178 MHz \citep{pilkington65}, the Sixth Cambridge Survey (6C) at 151MHz \citep{hales93}, the Parkes Radio Sources Catalogue (PKS) at 2.7GHz \citep{wright90} and at 4.8 GHz \citep{griffith94}, the MIT-Green Bank 5GHz Survey \citep{bennett86}, and the Australia Telescope 20GHz Survey Catalog (AT20G; \citealt{murphy10}). We found that NVSS is the optimal parent survey because of its similarity in the flux distribution and of its well characterized completeness\footnote{Despite using the NVSS as radio parent sample, a derivation of a luminosity function with another sample representative of the 6th Cambridge Survey of radio sources (6C survey) is also discussed in Section~\ref{sec:lco}, for comparison purposes.}.

\begin{figure}
\resizebox{\hsize}{!}{\includegraphics{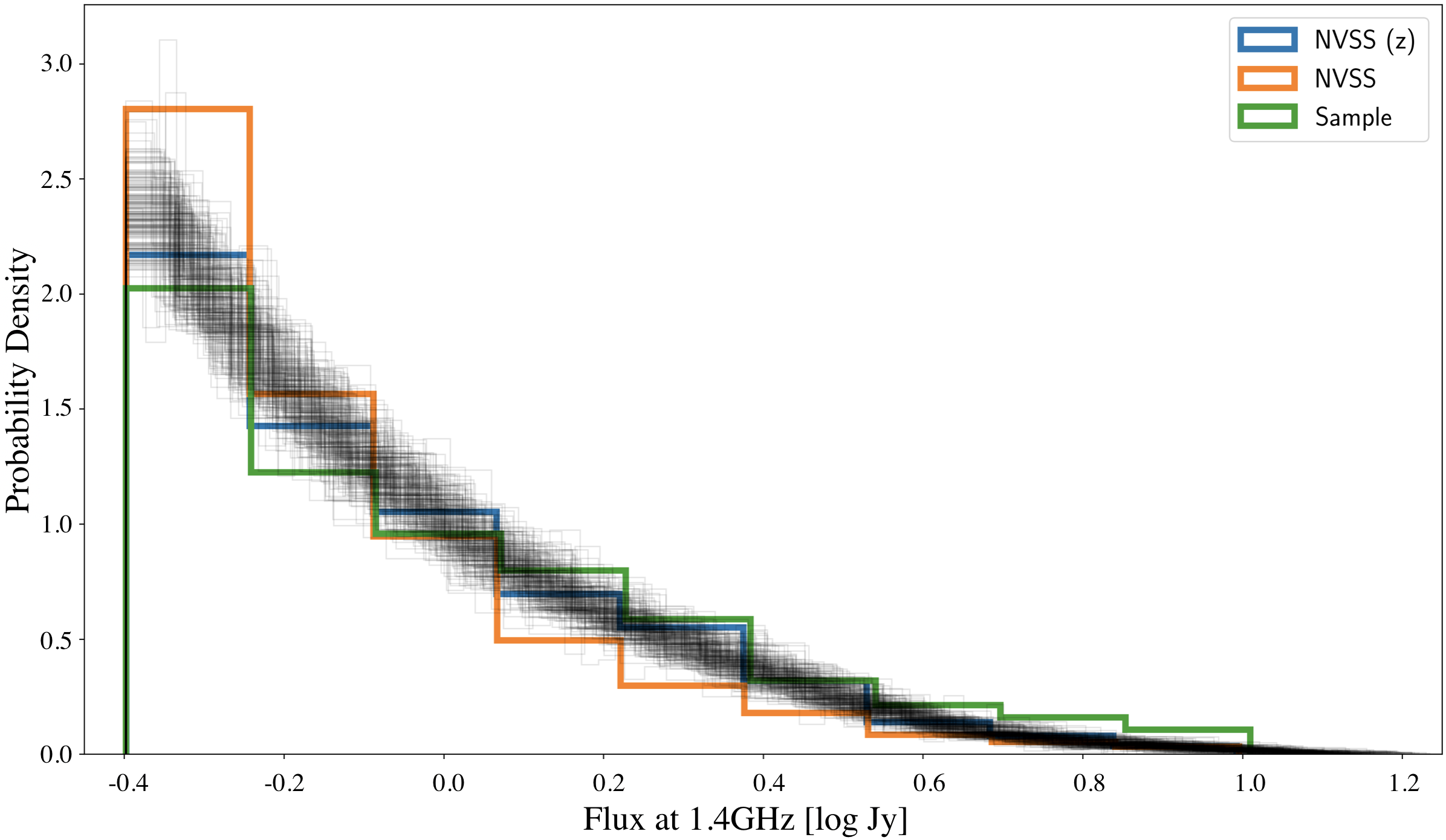}}
\caption{1.4 GHz flux distribution of NVSS and CO-ARC. The orange histogram shows the distribution for all NVSS parent sample, while the blue histogram is for the NVSS with redshift information. The thin black lines indicate the re-shuffled samples of NVSS with flux variations up to 10\%. The CO-ARC in green is well represented by both parent samples within the 10\% errors.}
\label{fig:flux_distributions}
\end{figure}

We then kept the largest possible sub-sample of the extended CO-ARC that is fully representative of the NVSS 1.4 GHz flux distribution. We used bootstrapping to iteratively and randomly remove more and more sources from our initial sample until we obtain a final sample consistent with the NVSS. Often, the convergence criterion of any two samples is checked with a Kolmogorov-Smirnov (KS) test: KS provides a significance level (p-value) based on the maximum distance between the two cumulative distributions (D-statistic), disproving the hypothesis that the two distributions are inconsistent. However, the p-value is rather arbitrary. Instead, we used our own precisely-defined D-statistic criterion, so that bootstrapping can reliably provide the maximum possible CO-ARC sub-sample that is compatible with the NVSS. For this definition, we simulated new, mock sets of NVSS data, starting from the initial fluxes in the catalogue and adding to them Gaussian random noise with dispersion $\sigma_f$. Then, we measured how the D-statistic between the original NVSS fluxes and any N-sized subsample of simulated NVSS fluxes changes as a function of N. The curves are shown in  Fig.~\ref{fig:convergence}, and they are calculated using the average D-statistic of 1000 implementations for each N. They indicate how close to the parent catalogue a child catalogue would be if its fluxes varied by 1$\sigma_f$ and 3$\sigma_f$, respectively, using a typical flux uncertainty of $\sigma_f$=10\%. As the sample size is getting smaller, the corresponding D-statistic grows due to dominance of small number statistics: at small enough sample sizes any two samples can resemble each other and thus the statistics deteriorate.

\begin{figure}
\resizebox{\hsize}{!}{\includegraphics{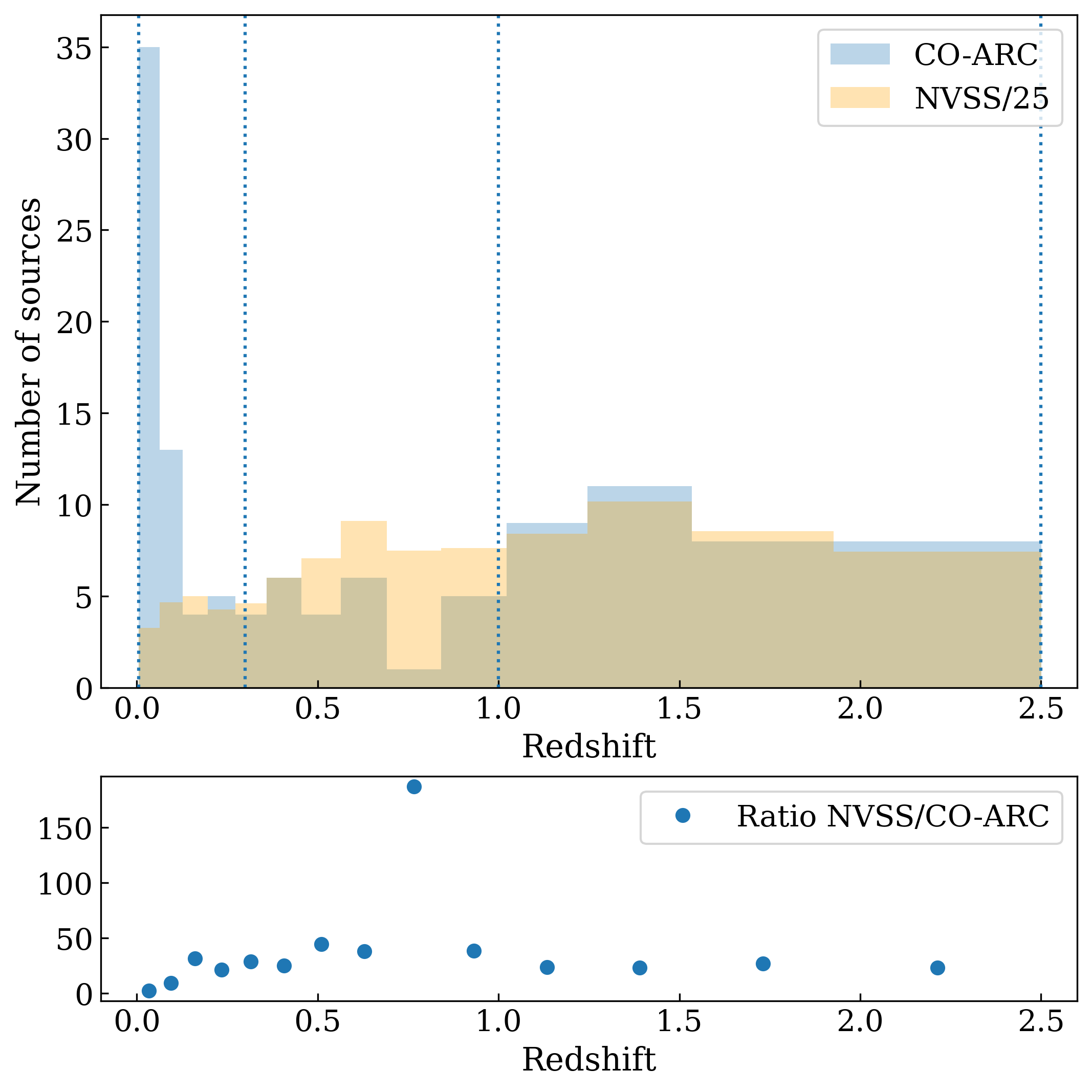}}
\caption{Redshift completeness of CO-ARC. {\it Top}: Number of sources as a function of $z$ in our final CO-ARC sample and in the NVSS parent catalogue (divided by 25 for comparison purposes). The NVSS distribution was scaled to a factor of 25 for displaying purposes. {\it Bottom:} Completeness correction so that our final sample reproduces the number of sources in the flux-limited NVSS with $z$. }
\label{fig:zbin}
\end{figure}

\begin{figure}
\resizebox{\hsize}{!}{\includegraphics{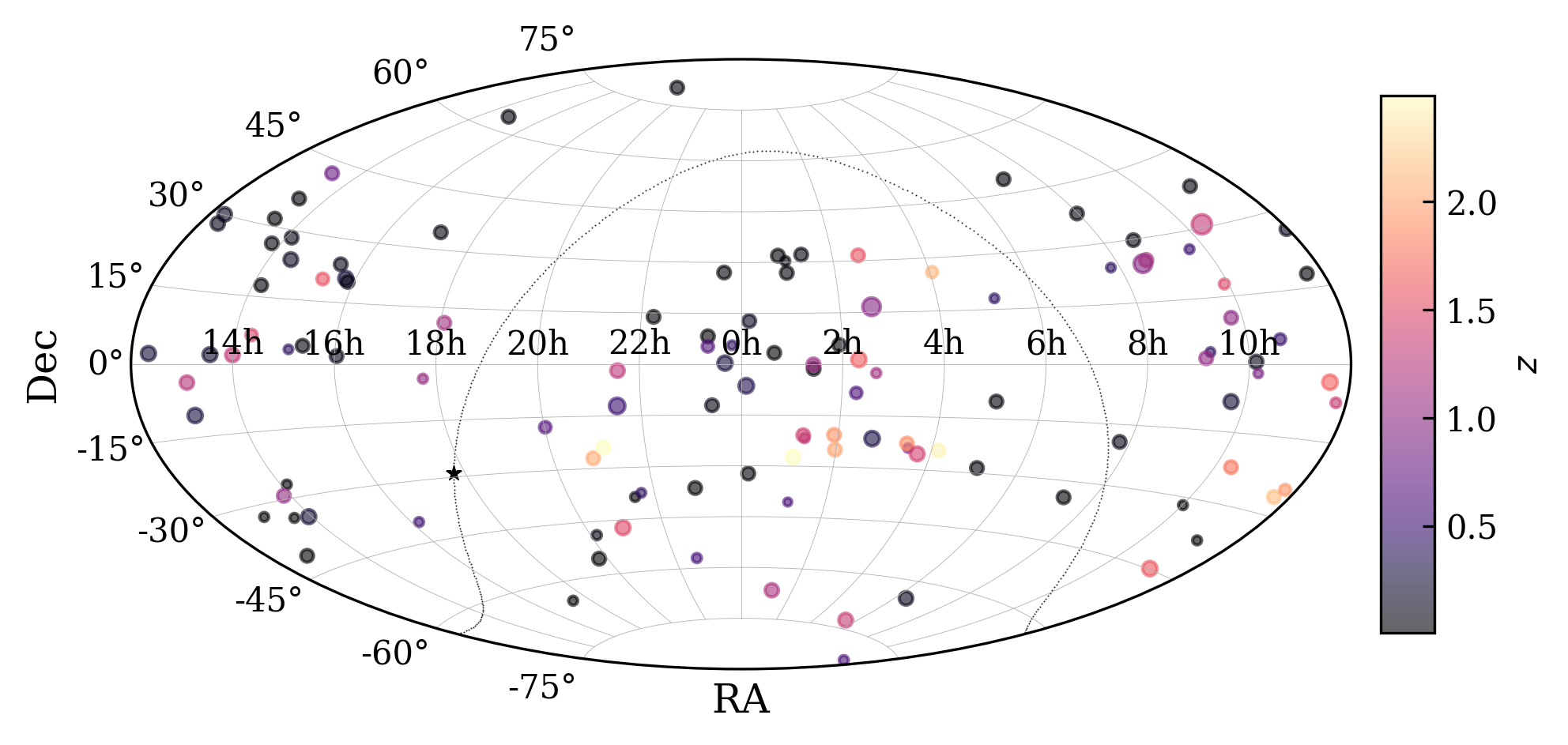}}
\caption{Distribution of the sample in the sky in the aitoff projection. The color code corresponds to the redshift and the sizes are according to the FWHP of the primary beam of the observations. The dotted curve and the star indicate the Galactic plane and center.}
\label{fig:sky}
\end{figure}

Having defined our theoretical convergence criterion, we proceeded with replacing the theorized child sample with the actual extended CO-ARC sample. Starting from the initial number of sources in the extended CO-ARC, we randomly drew consecutively smaller and smaller sub-samples and compared their D-statistic with the value established for the theorized NVSS sub-sample.
Our bootstrapping again indicates how the D-statistic changes as a function of N: it decreases with decreasing N as (mainly bright) sources are excluded from the sample, until it reaches a minimum and starts rising again due to small number statistics. The two samples become fully compatible (down to $\sim$ 1$\sigma_f$ or 10\% flux variation) for 120 sources. These numbers are presented for a flux cut of 0.4\,Jy. Looping over different flux thresholds and repeating the above procedure indicated that this choice led to the maximum possible sample size. The 1.4\,GHz flux distribution of our final sample and the flux-limited NVSS parent sample are shown in Figure~\ref{fig:flux_distributions}.

We also show the same analysis for the CO-ARC only using the ALMA calibrators (without including the literature sources, squares in Figure~\ref{fig:convergence}) and the CO-ARC with ALMA calibrators \textit{and} literature sources (circles) for the total sample and for the respective redshift bins, indicated by different color codes in Figure~\ref{fig:convergence}. The inclusion of literature sources helps to improve the statistics and to converge the CO-ARC sample to the theoretical D-statistic distribution of the shuffled NVSS.

It is noteworthy that our final sample is representative of the NVSS flux distribution also in each of the 3 individually examined $z$ ranges. We have chosen to use the ranges 0.005$<$z$<$0.3, 0.3$<$z$<$1,1$<$z$<$2.5, to study the evolution of the galaxies in comparable time steps of about 3.5-4.5\,Gyr. The CO-ARC sources of the $z$ slices are from 1$\sigma_f$ to 3$\sigma_f$ away from the NVSS parent catalogue at the same $z$. For the sake of completion, we note that this also holds separately for the sources from the ARC and the sources from the literature. The only exception comes from the few CO-ARC sources at intermediate $z$, where the construction of the luminosity function is unfeasible due to small number statistics. 

Having built a final sample that is representative of the flux-limited NVSS, we then calculated its completeness with respect to the parent survey number of sources in the sky as a function of $z$. For this purpose, we split both samples in even smaller $z$ steps (of 0.77\,Gyr) and present the comparison in Figure~\ref{fig:zbin}. The top panel shows the $z$ distribution of the two samples (with the flux-limited NVSS being divided by a factor of 25 for comparison purposes). It shows that the final CO-ARC is over-represented at $z$<0.2 and under-represented at z$\sim$0.8, but otherwise follows the parent catalogue's $z$ distribution. The completeness correction, i.e., the $z$-dependent number with which we need to multiply the CO-ARC galaxies to make them as frequent (in $z$ and in the sky) as the flux-limited NVSS, is shown in the bottom panel of Figure~\ref{fig:zbin} and is taken into account during the construction of the CO luminosity function in Section~\ref{sec:lco}.  There is one more correction to be taken into account in the luminosity function construction: the fact that not all NVSS sources have known $z$. On average, 1 on every 4 NVSS sources have $z$ information. This correction is, in principle, also $z$-dependent, but its behaviour with $z$ is unknown. We, thus, also multiply the number density of sources times 4, except for the local Universe (at $z$ $<$0.3), in which we assume the redshift acquisition to be complete.

\begin{table*}
\footnotesize
\centering  
\begin{tabular}{lccclllll}
\hline
   Galaxy  &    Line &   Beam size &  $\sigma_{\rm rms}$ &     S$_{\rm CO}\Delta v$ &    FWHM &   $v_{\rm res}$ &         log(MH$_2$) &      log(L'$\rm_{CO(1-0)}$) \\
   & & ("$\times$") & (mJy/beam) &  (Jy\kms) & (\kms) & (\kms) & (M$\rm_\odot$) & (K\kms pc$^2$) \\ 

\hline

 J1000-3139 &   CO(2-1)  &  0.97$\times$0.71 &       0.40 &  18.38$\pm$ 0.59 & 368.53 &  20.00 &    7.85$\pm$6.87 &    7.21$\pm$6.22 \\
 J1109-3732 &   CO(2-1)  &    0.6$\times$0.5 &       0.46 &   6.93$\pm$ 0.35 & 452.27 &  20.00 &    7.56$\pm$6.61 &    6.92$\pm$5.97 \\
 J1336-3357 &   CO(2-1)  &  0.63$\times$0.57 &       0.34 &   1.89$\pm$ 0.29 & 781.79 &  20.00 &    7.17$\pm$6.33 &    6.52$\pm$5.68 \\
 J1723-6500 &   CO(2-1)  &  0.27$\times$0.19 &       0.24 &  93.80$\pm$ 2.15 & 533.90 &  20.00 &    8.98$\pm$7.98 &    8.34$\pm$7.34 \\
 J1945-5520 &   CO(1-0)  &    1.8$\times$1.6 &       2.20 &         $<$17.99 & 300.00 & 100.00 &          $<$8.91 &          $<$8.28 \\
 J0057+3021 &   CO(2-1)  &   0.5$\times$0.32 &       0.18 &  13.52$\pm$ 0.20 & 582.32 &  20.00 &    8.27$\pm$7.29 &    7.63$\pm$6.64 \\
 J1301-3226 &   CO(2-1)  &  0.68$\times$0.65 &       0.31 &          $<$5.15 & 300.00 & 100.00 &          $<$7.87 &          $<$7.22 \\
 J2131-3837 &   CO(2-1)  &  0.67$\times$0.61 &       0.51 &   1.08$\pm$ 0.18 & 540.65 &  20.00 &    7.27$\pm$6.56 &    6.62$\pm$5.92 \\
 J1407-2701 &   CO(2-1)  &  0.69$\times$0.59 &       0.32 &   3.75$\pm$ 0.50 & 147.77 &  20.00 &    7.96$\pm$7.00 &    7.31$\pm$6.35 \\
 J2009-4849 &   CO(2-1)  &  0.64$\times$0.58 &       0.40 &   6.17$\pm$ 1.03 & 448.41 &  22.00 &    9.20$\pm$8.26 &    8.56$\pm$7.61 \\
 J1008+0029 &   CO(1-0)  &  2.21$\times$1.95 &       0.64 &          $<$0.74 & 300.00 & 100.00 &          $<$9.16 &          $<$8.52 \\
 J1221+2813 &   CO(1-0)  &  2.14$\times$1.39 &       0.07 &          $<$0.11 & 300.00 & 180.00 &          $<$8.40 &          $<$7.76 \\
 J0623-6436 &   CO(1-0)  &   0.79$\times$0.6 &       0.28 &   4.68$\pm$ 0.91 & 671.67 & 100.00 &   10.21$\pm$9.33 &    9.57$\pm$8.69 \\
 J1217+3007 &   CO(1-0)  &  2.56$\times$1.51 &       0.23 &          $<$0.21 & 300.00 & 100.00 &          $<$8.88 &          $<$8.24 \\
 J1427+2348 &   CO(1-0)  &  1.78$\times$1.32 &       0.27 &          $<$0.28 & 300.00 & 100.00 &          $<$9.18 &          $<$8.55 \\
 J1332+0200 &   CO(1-0)  &  2.74$\times$2.24 &       0.75 &          $<$0.39 & 300.00 & 100.00 &          $<$9.59 &          $<$8.95 \\
 J1356-3421 &   CO(1-0)  &   1.1$\times$0.88 &       0.28 &          $<$0.34 & 300.00 & 100.00 &          $<$9.56 &          $<$8.93 \\
 J0943-0819 &   CO(1-0)  &  4.78$\times$2.19 &       0.69 &          $<$0.36 & 300.00 & 100.00 &          $<$9.60 &          $<$8.96 \\
 J1220+0203 &   CO(1-0)  &  0.34$\times$0.26 &       0.36 &          $<$1.39 & 300.00 & 102.00 &         $<$10.24 &          $<$9.60 \\
 J1547+2052 &   CO(1-0)  &  0.23$\times$0.21 &       0.58 &          $<$2.83 & 300.00 & 100.00 &         $<$10.63 &          $<$9.99 \\
 J2341+0018 &   CO(1-0)  &   0.7$\times$0.57 &       0.55 &   4.16$\pm$ 0.56 & 350.00 &  20.00 &   10.84$\pm$9.93 &   10.20$\pm$9.29 \\
 J1305-1033 &   CO(1-0)  &  0.47$\times$0.42 &       0.14 &   0.67$\pm$ 0.17 & 173.67 & 100.00 &   10.06$\pm$9.25 &    9.42$\pm$8.61 \\
 J0242-2132 &   CO(1-0)  &  0.74$\times$0.63 &       0.30 &          $<$0.41 & 300.00 & 100.00 &          $<$9.95 &          $<$9.31 \\
 J0006-0623 &   CO(1-0)  &   2.27$\times$1.3 &      44.60 &         $<$23.17 & 300.00 & 100.00 &         $<$11.79 &         $<$11.15 \\
 J1505+0326 &   CO(3-2)  &  0.81$\times$0.44 &       0.81 &   3.28$\pm$ 0.26 & 376.56 &  20.00 &   10.15$\pm$9.31 &    9.50$\pm$8.66 \\
 J0748+2400 &   CO(3-2)  &  1.22$\times$0.99 &       0.61 &          $<$0.44 & 300.00 & 100.00 &          $<$9.28 &          $<$8.63 \\
 J0510+1800 &   CO(3-2)  &  0.65$\times$0.46 &       0.41 &          $<$0.26 & 300.00 &  20.00 &          $<$9.07 &          $<$8.42 \\
 J2349+0534 &   CO(3-2)  &  7.77$\times$5.09 &       3.13 &          $<$1.63 & 300.00 & 100.00 &          $<$9.87 &          $<$9.22 \\
 J2141-3729 &   CO(3-2)  &   1.4$\times$1.04 &       0.96 &   0.73$\pm$ 0.24 & 255.50 &  20.00 &    9.53$\pm$8.99 &    8.88$\pm$8.34 \\
 J0914+0245 &   CO(3-2)  &  0.05$\times$0.04 &       0.16 &          $<$3.34 & 300.00 & 100.00 &         $<$10.20 &          $<$9.55 \\
 J1038+0512 &   CO(2-1)  &  2.26$\times$1.94 &       0.95 &          $<$0.49 & 300.00 & 100.00 &          $<$9.80 &          $<$9.16 \\
 J0940+2603 &   CO(3-2)  &  0.11$\times$0.09 &       0.68 &          $<$3.58 & 300.00 &  60.00 &         $<$10.37 &          $<$9.71 \\
 J1610-3958 &   CO(3-2)  &  1.02$\times$0.81 &       0.53 &          $<$0.41 & 300.00 & 100.00 &          $<$9.46 &          $<$8.80 \\
 J2239-5701 &   CO(3-2)  &  6.96$\times$5.44 &       5.18 &          $<$2.69 & 300.00 & 100.00 &         $<$10.36 &          $<$9.71 \\
 J1058-8003 &   CO(3-2)  &  1.66$\times$1.09 &       0.89 &          $<$0.46 & 300.00 & 100.00 &          $<$9.61 &          $<$8.96 \\
 J0106-4034 &   CO(4-3)  &  7.07$\times$3.59 &       7.73 &          $<$4.02 & 300.00 & 100.00 &         $<$10.36 &          $<$9.66 \\
 J0217-0820 &   CO(2-1)  &  1.91$\times$1.46 &       0.58 &          $<$0.30 & 300.00 & 100.00 &          $<$9.81 &          $<$9.17 \\
 J2320+0513 &   CO(2-1)  &   1.2$\times$1.19 &       0.48 &          $<$0.25 & 300.00 & 100.00 &          $<$9.76 &          $<$9.12 \\
 J2000-1748 &   CO(2-1)  &   0.4$\times$0.27 &       0.36 &          $<$0.68 & 300.00 & 100.00 &         $<$10.23 &          $<$9.59 \\
 J1010-0200 &   CO(4-3)  &  1.17$\times$0.74 &       0.63 &          $<$0.37 & 300.00 & 100.00 &          $<$9.70 &          $<$9.00 \\
 J0329-2357 &   CO(4-3)  &  0.43$\times$0.33 &       0.54 &          $<$0.80 & 300.00 & 100.00 &         $<$10.03 &          $<$9.33 \\
 J0946+1017 &   CO(2-1)  &  0.39$\times$0.34 &       1.98 &          $<$2.92 & 300.00 & 100.00 &         $<$11.24 &         $<$10.59 \\
 J0909+0121 &   CO(2-1)  &  0.71$\times$0.66 &       1.12 &          $<$0.89 & 300.00 & 100.00 &         $<$10.74 &         $<$10.09 \\
 J1351-2912 &   CO(2-1)  &  2.32$\times$1.87 &       0.38 &          $<$0.20 & 300.00 & 100.00 &         $<$10.10 &          $<$9.45 \\
 J1743-0350 &   CO(4-3)  &  0.72$\times$0.55 &       0.06 &   2.26$\pm$ 0.04 & 145.99 &  80.00 &   10.62$\pm$9.65 &    9.92$\pm$8.95 \\
 J0125-0005 &   CO(2-1)  &  3.26$\times$2.52 &       0.42 &          $<$0.21 & 300.00 &  90.00 &         $<$10.15 &          $<$9.51 \\
 J0239-0234 &   CO(4-3)  &   0.4$\times$0.32 &       0.47 &          $<$0.70 & 300.00 & 100.00 &         $<$10.16 &          $<$9.46 \\
 J0837+2454 &   CO(2-1)  &   0.9$\times$0.79 &       0.66 &          $<$0.42 & 300.00 & 100.00 &         $<$10.49 &          $<$9.84 \\
 J0118-2141 &   CO(4-3)  &   0.6$\times$0.53 &       0.77 &          $<$0.72 & 300.00 & 100.00 &         $<$10.21 &          $<$9.51 \\
 J0112-6634 &   CO(2-1)  &  0.54$\times$0.47 &       0.61 &          $<$0.64 & 300.00 & 100.00 &         $<$10.72 &         $<$10.08 \\
 J1304-0346 &   CO(2-1)  &  0.71$\times$0.65 &       0.31 &          $<$0.23 & 300.00 & 100.00 &         $<$10.33 &          $<$9.68 \\
 J2134-0153 &   CO(2-1)  &  1.52$\times$1.25 &       0.47 &          $<$0.24 & 300.00 & 100.00 &         $<$10.37 &          $<$9.72 \\
 J1359+0159 &   CO(2-1)  &   0.7$\times$0.55 &       0.31 &          $<$0.26 & 300.00 & 100.00 &         $<$10.41 &          $<$9.77 \\
 J0529-7245 &   CO(2-1)  &  2.38$\times$2.16 &       0.89 &          $<$0.21 & 300.00 &  20.00 &         $<$10.33 &          $<$9.69 \\
 J1147-0724 &   CO(4-3)  &  0.55$\times$0.37 &       0.58 &          $<$0.66 & 300.00 & 100.00 &         $<$10.29 &          $<$9.59 \\
 J0343-2530 &   CO(2-1)  &  1.47$\times$1.19 &       0.38 &          $<$0.19 & 300.00 & 100.00 &         $<$10.35 &          $<$9.71 \\
 J1419+0628 &   CO(3-2)  &   0.39$\times$0.3 &       0.17 &   0.98$\pm$ 0.18 & 410.76 & 100.00 &   10.72$\pm$9.95 &   10.07$\pm$9.30 \\
 J0954+1743 &   CO(4-3)  &  0.87$\times$0.78 &       0.46 &   0.58$\pm$ 0.09 & 193.30 &  20.00 &   10.31$\pm$9.61 &    9.61$\pm$8.91 \\
 J2056-4714 &   CO(2-1)  &  2.23$\times$1.77 &       0.23 &          $<$0.12 & 300.00 & 100.00 &         $<$10.18 &          $<$9.54 \\
 J1520+2016 &   CO(3-2)  &  0.47$\times$0.32 &       0.13 &          $<$0.17 & 300.00 & 100.00 &         $<$10.03 &          $<$9.38 \\
 J1107-4449 &   CO(2-1)  &   0.97$\times$0.8 &       0.91 &   4.98$\pm$ 0.15 &  43.74 &  20.00 &  11.86$\pm$10.90 &  11.21$\pm$10.26 \\
 J0219+0120 &   CO(2-1)  &  0.86$\times$0.65 &       0.66 &          $<$0.45 & 300.00 & 100.00 &         $<$10.82 &         $<$10.18 \\
 J1136-0330 &   CO(2-1)  &  2.49$\times$2.19 &       0.55 &          $<$0.29 & 300.00 & 100.00 &         $<$10.64 &         $<$10.00 \\
 J1146-2447 &   CO(4-3)  &   2.3$\times$1.47 &       0.49 &          $<$0.25 & 300.00 & 100.00 &         $<$10.17 &          $<$9.47 \\
 J0403+2600 &   CO(4-3)  &  0.06$\times$0.04 &       0.11 &          $<$1.07 & 300.00 & 100.00 &         $<$10.86 &         $<$10.16 \\
 J0106-2718 &   CO(3-2)  &  1.43$\times$1.09 &       0.59 &          $<$0.31 & 300.00 & 100.00 &         $<$10.64 &          $<$9.99 \\
 
\hline
\end{tabular}
\caption{CO properties of the 66 CO-ARC sample sources. Note that, even for targets with available multiple-J CO observations, only the properties of lowest-J CO transitions are listed here and used for the luminosity function computations.}\label{tab:detections}
\end{table*}

In Figure \ref{fig:sky}, we show the distribution of the 120 sources of our survey in the sky plane projection. The symbol sizes are indicative of the full width at half power (FWHP) of the primary beam of the  observations. The total survey area covered by our sample was estimated by summing the FWHP of all individuals observations, taking into account the dependency of the FWHP on the frequency of each observed CO line. This is feasible as the sources are distributed randomly in the sky with almost no overlap between them.

Our final sample, consisting of 120 sources and being representative of the NVSS down to 0.4 Jy, contains 66 sources from the CO-ARC plus 54 literature sources. Details on the CO-ARC sources are given in Tables~B.1 and 1
, including  the spectral line of interest per galaxy, the ID of the ALMA observations that we found and reduced per line, as well as the redshift and its source of origin per galaxy.  For the bibliographic sources, all pertinent information is summarized in Table~\ref{tab:biblio} in the Appendix.

\section{ALMA Observations}
\label{sec:obs}

\subsection{Data reduction}

The ALMA observations that were analyzed as part of the CO-ARC survey were carried out between Cycle 0 and Cycle 5. The sources were observed as flux, phase, bandpass, or polarization calibrators. The relevant properties of each observing run (e.g.\,scan intent, on-source integration time, CO transition, etc.\,) are reported in Table~\ref{tab:proj}. 

The data of the sources observed as flux, phase, and polarization calibrators were reduced using the Common Astronomy Software Application \citep[{\sc CASA};][]{casa} pipeline, which automatically processes the data by performing standard calibration and flagging. Various pipeline versions were used, according to those adopted for the corresponding archival data reduction. When one of our targets was observed in more than one project, once calibrated separately according to the appropriate  {\sc CASA} version, the visibilities were then combined using the {\sc CASA} task {\tt concat}. In a minority of cases the projects were observed in the first cycles (0, 1 and early 2) and were calibrated using casa version <=4.2.2, which computes the visibility weights in a different way and therefore they had to be corrected to the same definition using the {\sc casa} task {\tt statwt}\footnote{for more details, see \url{https://casaguides.nrao.edu/index.php?title=DataWeightsAndCombination}}. 

Bright quasars were typically used as flux calibrators, resulting in a standard 10\%  flux calibration uncertainty.
A different procedure was required for sample sources observed as bandpass calibrators, as any line emission in the source would be flattened by the default pipeline corrections. This is because the bandpass calibration corrects the observed visibilities for  frequency-dependent amplitude and phase corruptions. Such corrections are calculated using as model the bandpass calibrator, which is typically a bright quasar (i.e.\,a point-like source in the sky). Due to the Fourier Transform properties, a point-like sky brightness distribution ideally has flat and constant amplitudes in the visibility plane. The bandpass calibration calculates the corrections needed to make the data as similar as possible to such ideal model. Therefore, any line emission in the spectrum of the bandpass calibrator would be treated as instrumental fluctuation and thus flattened.  In such cases, the data were then manually calibrated using customised data reduction scripts, which were modified so that the bandpass calibrator was substituted by the phase (or flux) calibrator in the relevant calibration steps. A standard data reduction process was then carried out also in these cases. 

\subsection{Imaging}
Where detected, the CO line emission was identified and isolated in the $uv$-plane using the {\sc casa} task {\tt uvcontsub}. This latter forms a continuum model from linear fits in frequency to line-free channels, and then subtracts this model from the visibilities.

Deconvolution and imaging were performed using the {\sc CASA} task {\tt tclean} with Briggs weighting and two robust parameters: 0.5 and 2. The former was chosen as it provides a good trade-off between angular resolution and signal-to-noise ratio (S/N); the latter the best sensitivity. For each type of weighting, we produced three data cubes with different channel widths: 20, 60 and 100~km~s$^{-1}$. For low spectral resolution observations in which the raw channel width was larger than 20 or 60~km~s$^{-1}$, we adopted the lowest possible channel binning. The continuum-subtracted dirty cubes were iteratively cleaned in regions with line emission (first identified through auto-masking) down to selected  
thresholds, and then primary-beam corrected.

\begin{figure}
\resizebox{\hsize}{!}{\includegraphics{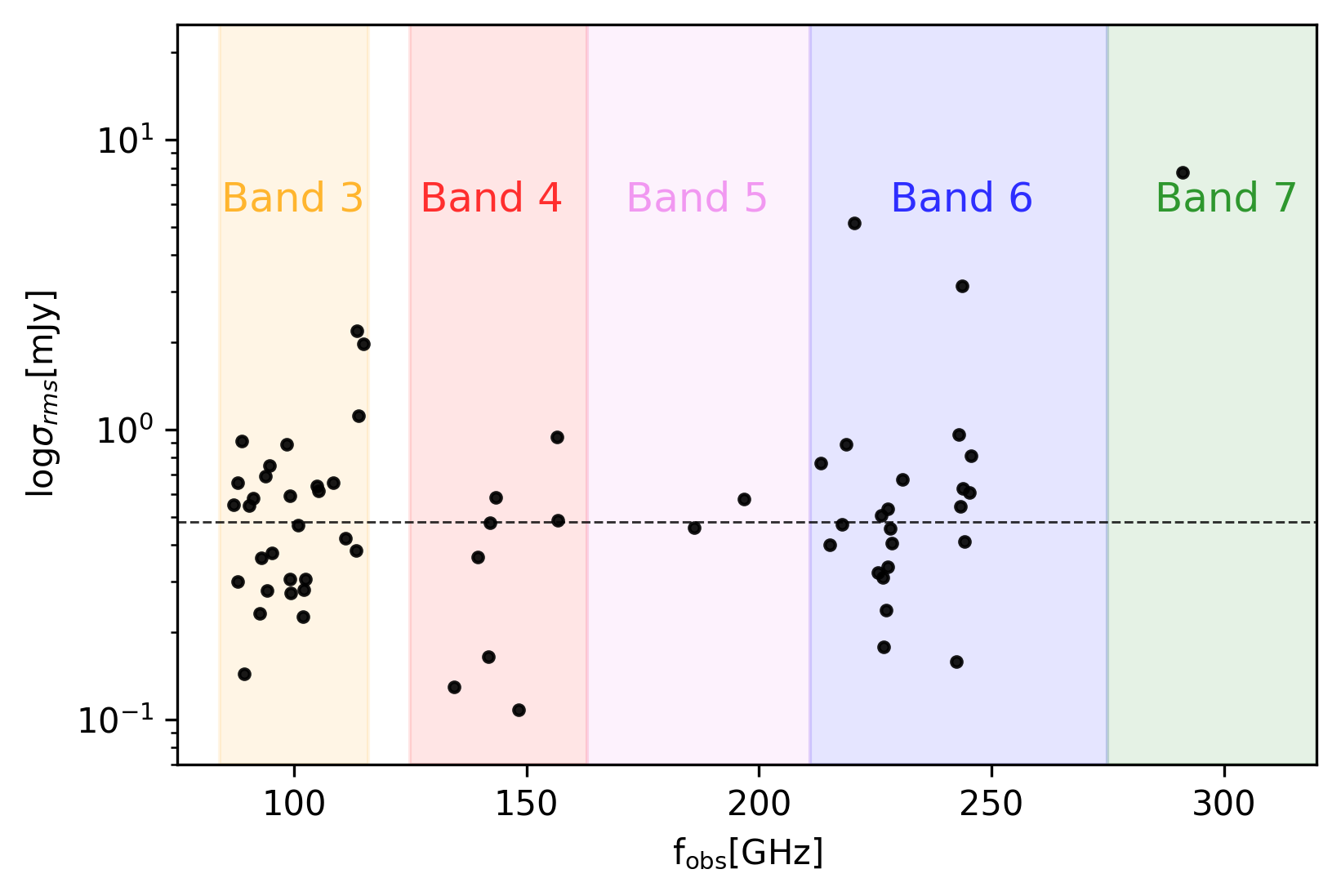}}
\caption{Distribution of achieved rms sensitivities per ALMA band. The black circles represent the rms noise levels (in log scale and mJy units) achieve for the emission line datacubes of the ALMA calibrators sample in function of the respective observed frequencies (in GHz) or ALMA bands highlighted in colors. The dashed line indicates the median value of 0.482\,mJy for the 66 sources drawn from the CO-ARC.}
\label{fig:noise}
\end{figure}

The achieved synthesised beams and root mean square (rms) noise levels of both detections and non-detections are listed in Table~\ref{tab:detections}. The median rms noise obtained for the emission line data cubes of ALMA calibrators is 0.48\,mJy and some scatter is seen along this value, however, overall the rms noises per ALMA band are fairly homogenous and below 1\,mJy for most of the targets, as we can see in Figure~\ref{fig:noise} the distribution of $\sigma_{\rm rms}$ per frequency (or per ALMA band). We note that even though some of the CO-ARC sources have been presented in the literature before, 
we opted to re-image them, so that our archival survey has homogeneous imaging procedure. 

\section{Results}
\label{sec:results}

\subsection{Moment maps and spectra of CO-ARC detections}\label{mommaps}

Overall, we report the detection of CO emission in 17 out of the 66 CO-ARC sources listed in Table~\ref{tab:detections}. A marginal detection for J2341+0018 (that  most likely reflects a potential flux loss/bad calibration of its data). The remaining 49 analyzed sources are undetected in CO. Together with the 25 CO detections in the 54 literature sources listed in Table~\ref{tab:biblio}, we find that 35\% of radio galaxies in the extended CO-ARC sample contain detectable gas reservoirs. The fraction differs for local and high-$z$ sources, varying from 31/60 sources in the local Universe to 2/23, and 9/37 sources at intermediate and high-$z$, respectively.

Integrated intensity (moment 0), mean line-of-sight velocity (moment 1), and velocity dispersion (moment 2) maps of the CO-ARC detections were created from the continuum-subtracted, cleaned data cubes, keeping pixels down to 3$\sigma$ noise levels. The resulting maps are shown in Figure~\ref{fig:detsam} and in the Appendix in Figure~\ref{fig:detall}. 

\begin{figure*}
\centering
\includegraphics[width=17cm]{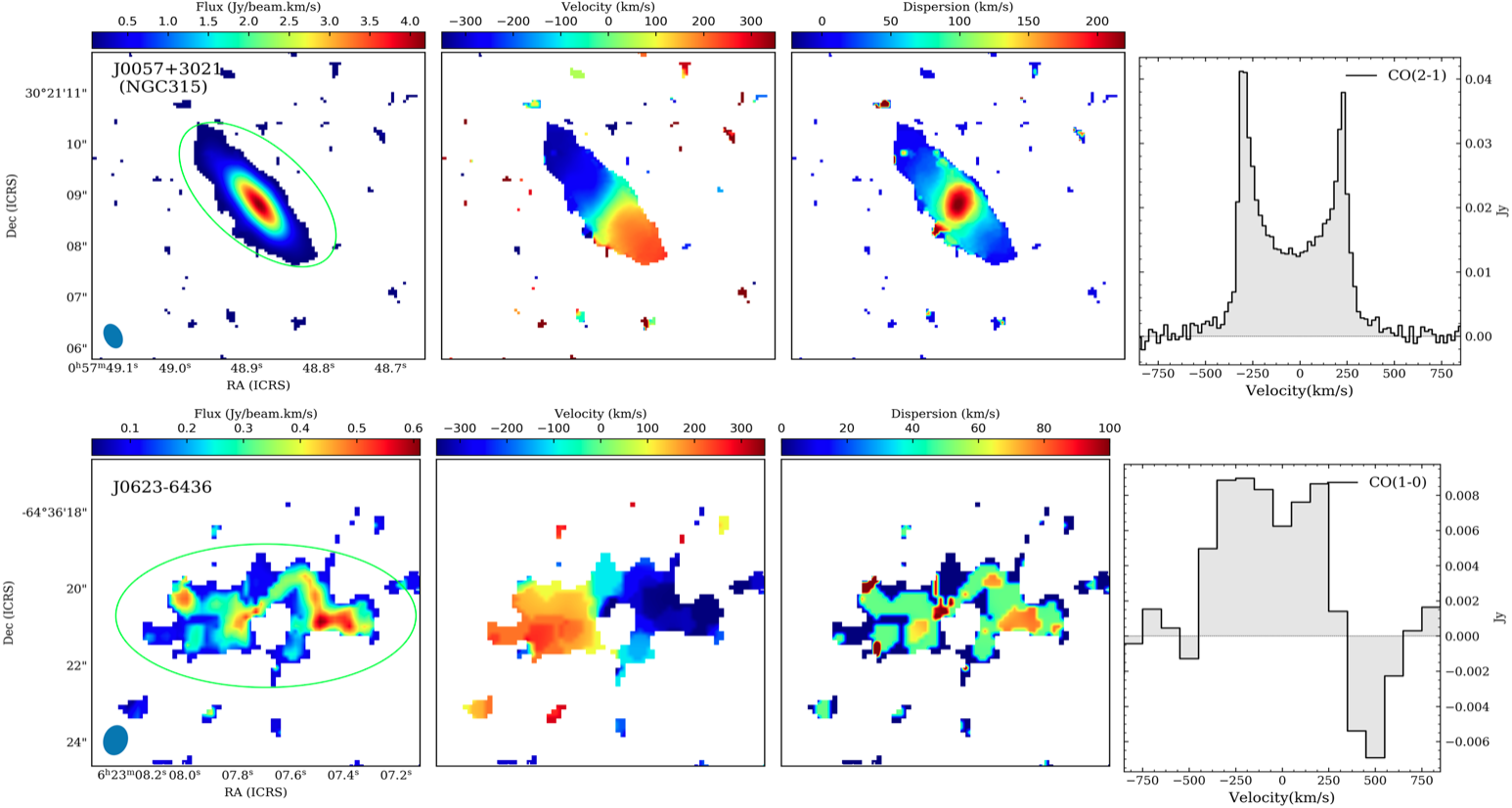}
\caption{Example of integrated intensity (moment 0; left panels), mean line-of-sight velocity (moment 1; middle-left panels), velocity dispersion (moment 2; middle-right) maps and spectra (right panels) of 2 detections in the CO-ARC sample. The moments maps for all the 17 detections are shown in the Appendix in Figure~\ref{fig:detall}. The synthesized beam is shown as a blue ellipse in the bottom-left corner of each moment 0 map. The bar to the top of each moment map shows the color scales (in Jy~beam$^{-1}$~km~s$^{-1}$ and km~s$^{-1}$ for moment 0 and moment 1/2 maps, respectively). East is to the left and north to the top. The observed CO transition is indicated in the top-right corner of each spectral profile. The aperture within which the illustrated spectra have been extracted are overlain in green in each moment 0 maps.}
\label{fig:detsam}
\end{figure*}

Based on the integrated intensity maps (left panels of Fig.\,\ref{fig:detsam} and Fig.\,\ref{fig:detall}), spatially-resolved extended gas distributions are observed in six targets (NGC\,315, IC\,4374, NGC\,3100, J0623-6436, NGC\,6328, and J2009-4349), most of which presenting disc- or ring-like structures. Marginally-resolved or unresolved gas structures are instead observed in all the other cases. The mean line-of-sight velocity maps (Fig.\,\ref{fig:detsam} and Fig.\,\ref{fig:detall}, middle-left panels) show clear evidence of gas rotation in eight cases, that are NGC\,315, IC\,4296, NGC\,3557, NGC\,3100, J0623-6436, NGC\,6328, J2009-4849, and NGC\,7075. Among these, clear kinematic distortions (i.e.\,s-shaped iso-velocity contours) are observed in IC\,4296, NGC\,3100, NGC\,6328. Such disturbances usually trace the presence of unrelaxed sub-structures in the gas distribution, possibly caused by either warps or non-circular motions, or a combination of both.

Accurate 3D kinematic modelling can help differentiating between the two. This has been carried out in details for IC\,4296 and NGC\,3100 by \citet{ruffa19a,ruffa22}. In the former case, such an analysis led to the conclusion that the presence of a position angle warp best reproduces the observed kinematic asymmetries. The additional presence of non-circular motions, however, cannot be fully excluded nor fully established due to the fact that the gas distribution is only marginally resolved in IC\,4296. The features observed in NGC\,3100 are instead found to be the result of a more complex kinematics, including the presence of both a position angle and inclination warp and non-circular motions. These latter have been recently partly identified into a mild jet-induced outflow with $\dot{M}$=0.12~M$_\odot$~yr$^{-1}$ \citep{ruffa22}.

Another case of full kinematic modeling was that of NGC\,6328, which has been carried out in the context of the CO-ARC project and can be found in separate papers (\citealt{michalis21}; Papachristou et al. 2022 in prep.). In short, our detailed analysis suggests the presence of a jet-induced of outflow as likely explanation for the kinematics observed in NGC\,6328, in addition to a highly warped gas distribution. A massive molecular outflow with $\dot{M}$=2300~M$_\odot$~yr$^{-1}$ has been also recently reported by \citet{vayner17} in the galaxy 3C\,298.

The line-of-sight gas velocity dispersion ($\sigma_{\rm gas}$; Figure~\ref{fig:detsam} and Figure~\ref{fig:detall}, middle-right panels) varies significantly among different detections and across the gas sky distribution observed in each source. In most of the detections with disc- or ring-like shapes (e.g.\,NGC\,315, J1505+0326, IC\,4296, NGC\,3557, NGC\,3100, etc.\,), the gas is observed to be dynamically cold towards outskirts of CO distribution, with $\sigma_{\rm gas}\sim10-20$~km~s$^{-1}$ in agreement with the typical ranges expected in unperturbed cold gas clouds \citep[e.g.][]{Voort18}. $\sigma_{\rm gas}$ then progressively increases towards the central gas regions, with the largest values ($\geq50$~km~s$^{-1}$) typically observed around the location of the line peak(s). In first instance, such large values would imply the presence of turbulence within the gas clouds. Observed line-of-sight velocity dispersions, however, can be significantly overestimated due to observational effects. Among these, the limited spatial resolution and/or low detection significance can introduce severe beam smearing (i.e.\,contamination from partially resolved velocity gradients within the host galaxy), which usually dominates in areas of large velocity gradients (such as the central regions of galaxies). Full kinematic modelling is again the only way to derive reliable estimates of the intrinsic gas velocity dispersion: this has been carried out in details for NGC\,315 by \citet{boizelle21} and for IC\,4296, NGC\,3557, NGC\,3100, and NGC\,7075 by \citet{ruffa19a,ruffa22}.

The spatially-integrated CO spectra of the 17 CO-ARC detections are shown in the right panels of Figure~\ref{fig:detall}. These have been extracted from the cleaned data cubes within circular or elliptical apertures enclosing all the observed CO emission (as illustrated in the left panels of Fig.\,~\ref{fig:detsam} and Fig.\,~\ref{fig:detall}). All of the eight above-mentioned sources showing clear signs of rotation in their moment 1 maps also exhibit the classic double-horned spectral shape expected from a rotating disc. In addition, in IC\,4296 a strong absorption feature is observed. This has been analyzed in detail by \citet{ruffa19b} and mainly interpreted as absorption against a bright core continuum. In all the other detections, the spectral profiles show Gaussian-like, centrally peaked shapes. The integrated CO flux of each target, $S_{\rm CO}\Delta v$, has been estimated from the obtained spectral profiles and, together with the line full width at half-maximum (FWHM), is listed in Table~\ref{tab:detections}. 

For non-detections, upper limits were calculated in a customary Poissonian way multiplying 3$\sigma_{\rm rms}$ (i.e.\,three times the root mean square noise level determined from line-free channels of each data cube) by the square root of the number of channels contained in an assumed line of FWHM=300\kms\ . We used the $\sigma_{\rm rms}$ value of the data cubes made with robustness parameter equal to 2 and 100\kms channel width. This is valid if we consider that the whole CO emission is concentrated within the synthesized beam, which is likely the case for the most distant sources. For nearby targets, to take into account possibly extended emission, we assumed a typical galaxy size of 10\,kpc and corrected the previous noise by multiplying it by the square root of the number of synthesized beams inside the assumed galaxy size.

\subsection{Luminosity and Molecular mass}

A common way to express the CO luminosity is via $L^\prime_{\rm CO}$ in units of $\rm K~km~s^{-1}~pc^{2}$, which is proportional to the surface brightness temperature, T$_{\rm B}$. Its advantage comes from its definition: since it originates from the Rayleigh-Jeans flux multiplied by the beam area, the frequency squared in the Rayleigh-Jeans law gets cancelled out with $1/\nu^2$ of the beam area. 
More specifically, the equation of \citet{sol05} provides $L^\prime_{\rm CO}$
from the integrated fluxes $\rm S_{CO}\Delta V$ as:
\begin{equation}
{L^\prime_{\rm CO}}(K km s^{-1} pc^2) = 3.25\times10^7 \frac{S_{\rm CO}\Delta V}{(1+z)^3}  \left( \frac{D_{\rm L}}{\nu_{\rm obs}} \right)^2, 
\end{equation}
where $\nu_{\rm obs}$ is the observed frequency of the CO line, in GHz, $D_{\rm L}$ is the luminosity distance of the galaxy, in Mpc, $z$ is the redshift and $\rm S_{\rm CO}\Delta V$ is the integrated flux in Jy\kms. The mid-J  CO transitions can be converted to $L^\prime_{\rm CO(1-0)}$ adopting a line ratio, called r$_{\rm J\rightarrow1}=\frac{L^\prime_{\rm CO}(J\rightarrow {J-1} )}{L^\prime_{\rm CO}(1\rightarrow0)}$. For thermalized, optically thick gas emission, r$_{\rm J\rightarrow1}$=1. We adopted the r$_{\rm J\rightarrow1}$ values reported by \citet{carilli13}, r$_{\rm 2\rightarrow1}$=0.99, r$_{\rm 3\rightarrow1}$=0.97 and r$_{\rm 4\rightarrow1}$=0.87, for typical gas excitation conditions of QSOs. The values derived for the CO-ARC sources are in the range 6.5$<$log$L^\prime_{\rm CO(1-0)}<$11.1, while the values reported for sources in the literature sample are 5.4$<$log$L^\prime_{\rm CO(1-0)}<$10.7.

\begin{figure}
\resizebox{\hsize}{!}{\includegraphics{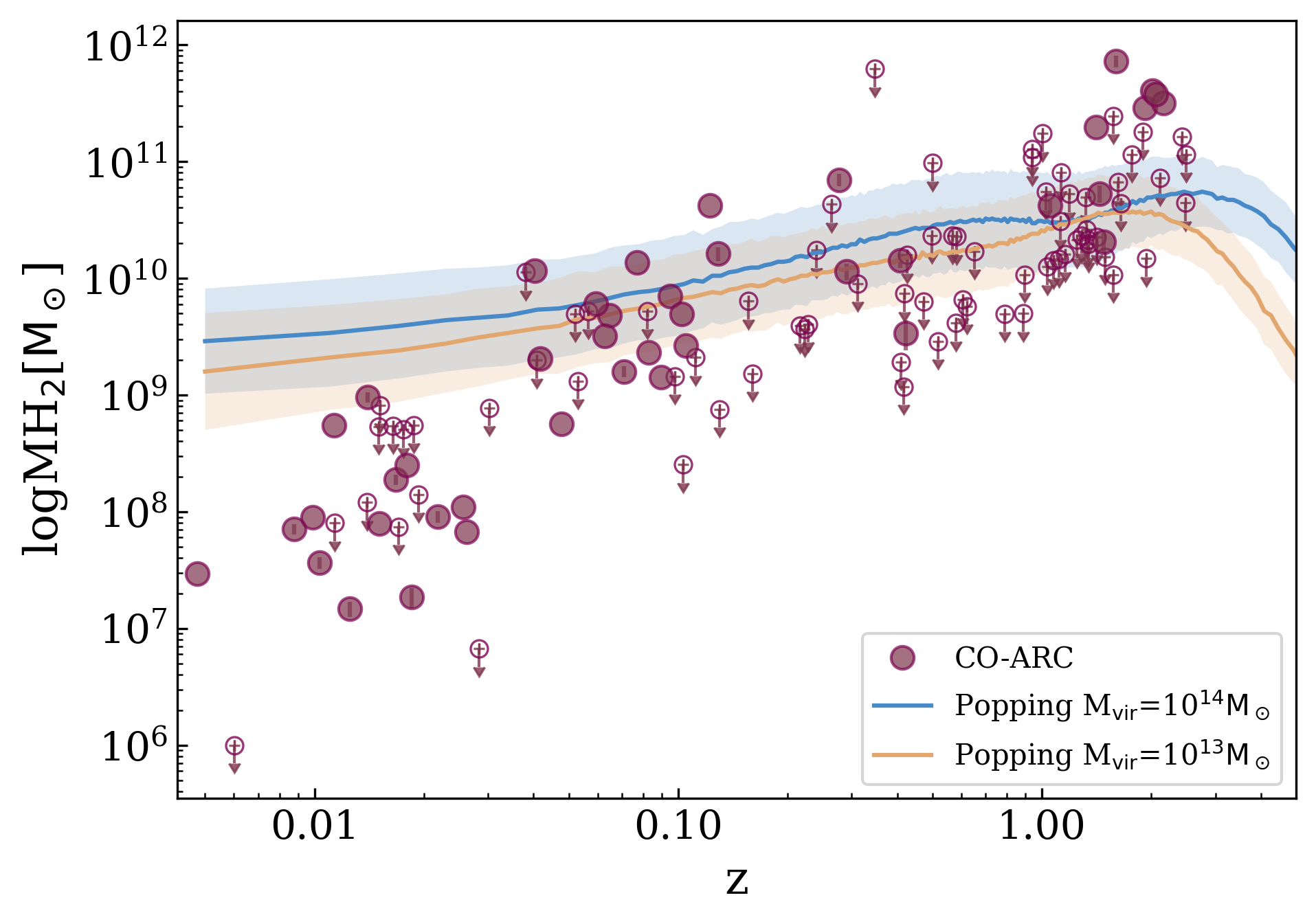}}
\caption{Molecular masses derived for the CO-ARC sample. Detection are shown with filled circles and upper limits are indicated with open circles and arrows. For comparison, we show the predicted evolution of the molecular gas content with redshift from semi-empirical models of normal star-forming galaxies with halo masses of $M_{\rm vir}=10^{13}$ and $M_{\rm vir}=10^{14}$ \citep{popping15}.} 
\label{fig:mgas}
\end{figure}

 Molecular masses can be calculated from ${L^\prime_{\rm CO(1-0)}}$  using the CO-to-H$_2$ conversion factor $M_{\rm H_2}=\alpha_{\rm CO} {L^\prime_{\rm CO(1-0)}}$.
 We adopted the  standard Galactic CO-to-H2, conversion factor of $\alpha_{\rm CO}=4.36M_\odot$ ($\rm K~km~s^{-1}~pc^{2}$)$^{-1}$  \citep{tacconi13,bolatto13} for consistency throughout the sample. Based on these assumptions, we derived molecular masses in the range of 1.5$\times10^{7}<M_{\rm{H_2}}<7.2\times10^{11} M_\odot$ for the 66 CO-ARC radio galaxies. For the 54 literature-based galaxies, including detections and upper limits, the reported $M_{\rm H_2}$ values vary from 1$\times10^{6}$ to 4$\times10^{11}M_\odot$, converting to the same $\alpha_{\rm CO}=4.36$ used in this work for consistency.

Figure~\ref{fig:mgas} illustrates the molecular masses as a function of $z$ for all the sources analyzed in this work. We compare our mass measurements from semi-empirical predictions of the typical H$_2$ gas content of galaxy halos from \citet{popping15}, with halo masses of $M_{\rm vir}=10^{13}$ and $M_{\rm vir}=10^{14}$. Interestingly, we find that even though at the very local Universe all radio galaxies have molecular gas contents significantly (1-2 orders of magnitude) below those predicted for typical galaxy halo values, the picture changes at high-$z$. At $z$ $>$1, several radio galaxies have molecular gas reservoirs that are very similar to (and even higher than) those of the typical galaxy halo contents. The detected molecular gas reservoirs of radio galaxies at high-$z$ ranges from 5$\times$10$^9$ to $\sim$10$^{12}$ $M_\odot$, while at low $z$ it ranges from 10$^7$ to 10$^{10}M_\odot$. Still, at all $z$, the majority of galaxies have small (undetected) reservoirs than that of typical galaxy halos. 

\begin{figure*}
\centering
\begin{minipage}{.47\textwidth}
  \centering
  \includegraphics[width=\linewidth]{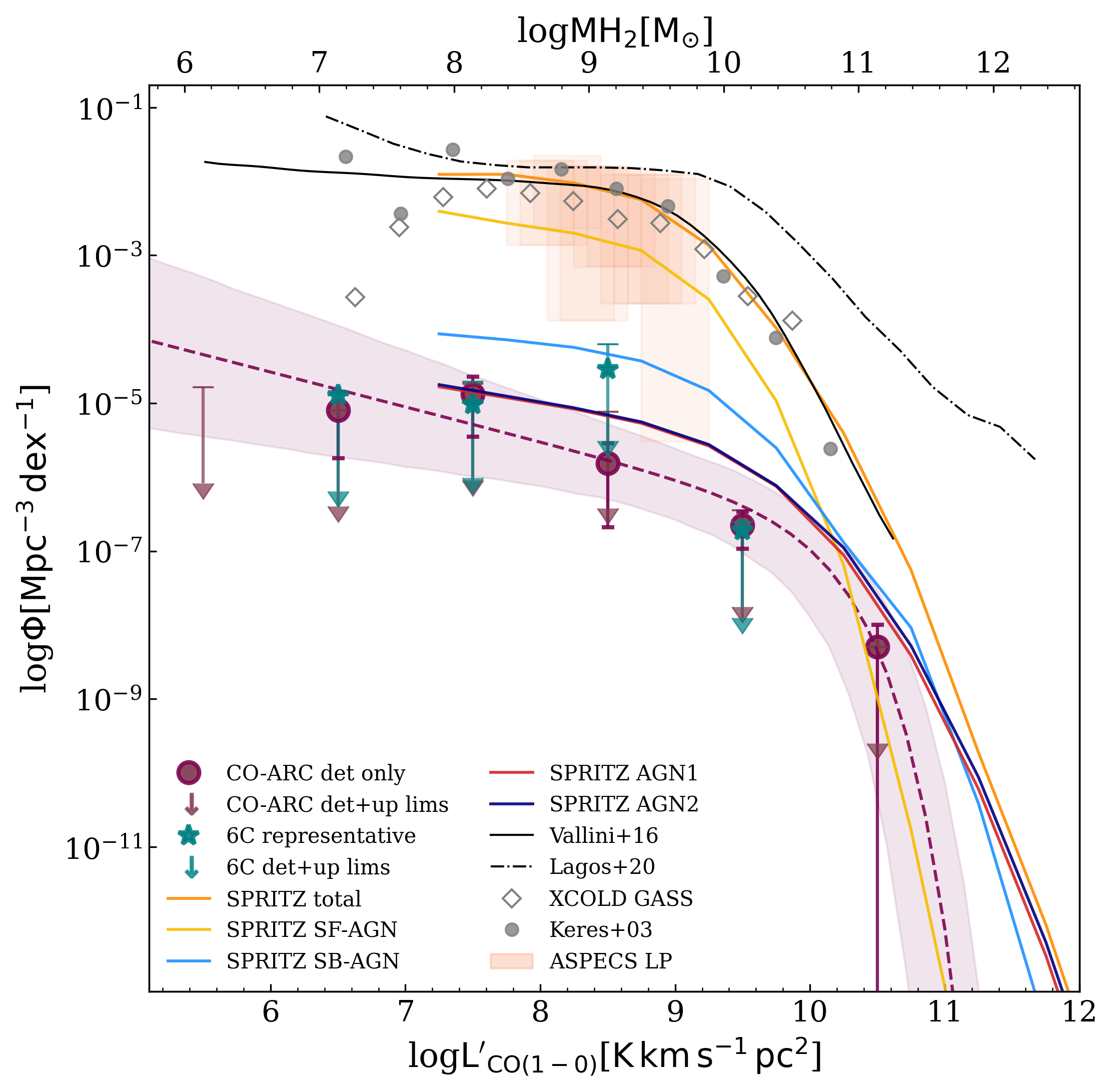}
\end{minipage}
\begin{minipage}{.47\textwidth}
  \centering
  \includegraphics[width=\linewidth]{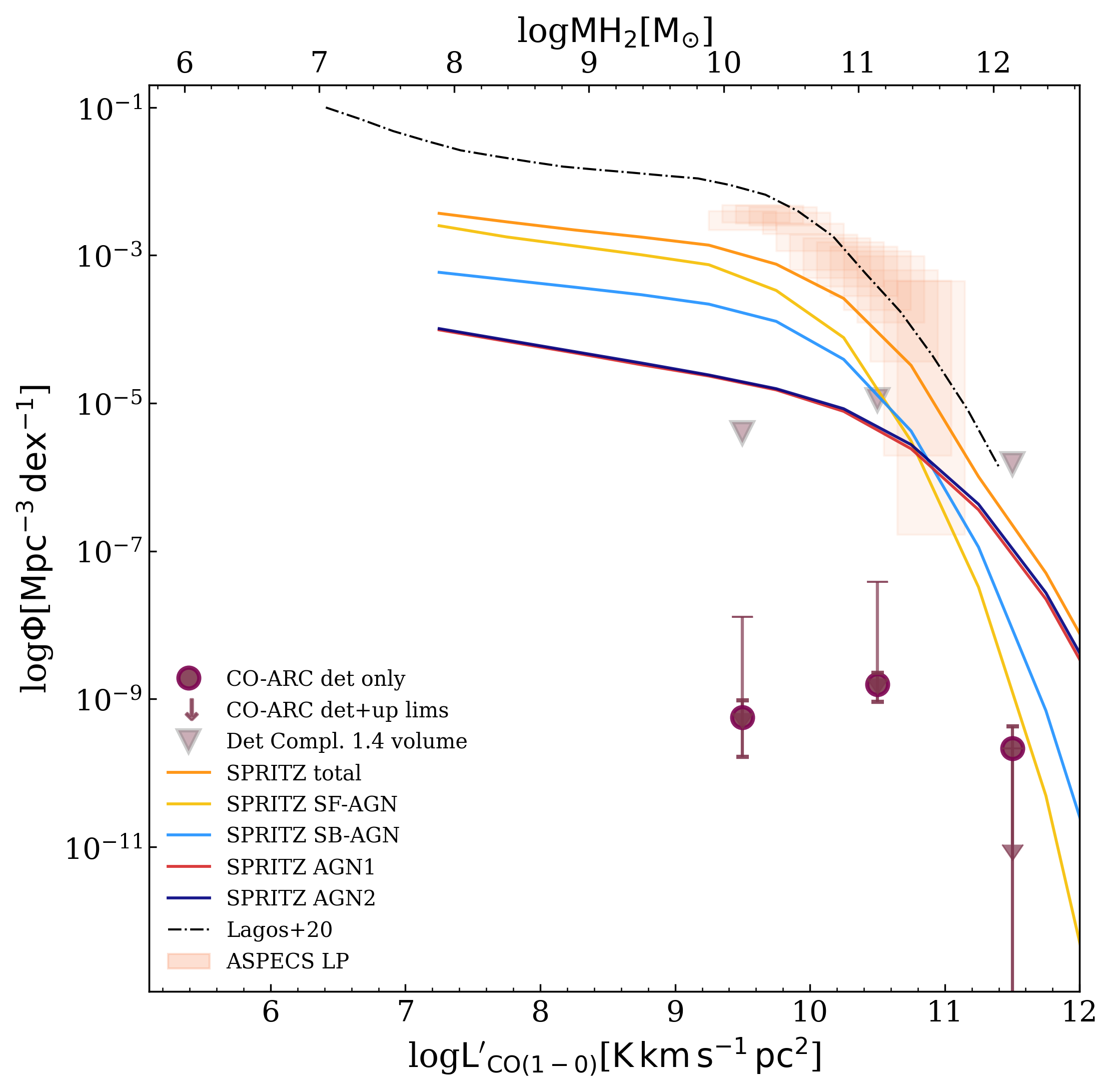}
\end{minipage}
\caption{Luminosity functions for $0.005<z<$0.3 in the left panel and for 1$<z<$2.5 in the right panel. For the local mass and luminosity functions we compare our findings with those of different surveys: the XCOLD GASS \citep[][open diamonds]{saintonge17,fletcher21}, as the gray dots the FIR-selected CO survey of \citet{keres03} from the IRAS bright galaxy sample, the empirical CO luminosity function derived from the \textit{Herschel} IR luminosity \citep[][black line]{vallini16} and the light orange boxes are results from the ASPECS LP \citep{decarli19}. The  {\sc shark} semi-analytic models  by \citet{lagos20} for the CO luminosity function of sub-mm selected galaxies are shown in dotted lines. Predictions from the SPRITZ simulation \citep{spritz} are show in orange for all galaxies, and for the different AGN populations in yellow for SF-AGN, in light blue for SB-AGN, and for AGN1 and AGN2 are indicated in red and dark blue, respectively.}
\label{fig:lco} 
\end{figure*}

\subsection{CO Luminosity Function}\label{sec:lco}

Previous studies reported luminosity functions for  normal and starburst galaxies in the local Universe \citep{keres03, obreschkow09, saintonge17}, however this is the first time that the CO luminosity function of radio galaxies is presented. To construct the luminosity function, we adopted the 1/$V_{\rm max}$ method of \citep{schmidt68}, which provides the number density for each luminosity bin:
\begin{equation}
\Phi (Mpc^{-3} dex^{-1}) = \frac{1}{\Delta\log L} \sum_{i} \frac{\omega_i}{ V_{max,i} },
\end{equation}
where $\Delta\log L$ is the luminosity bin size (common in logarithmic scale), $V_{max}$ is the maximum comoving volume that any galaxy $i$ could be observable given its measured 1.4 GHz flux and the survey limit and $\omega_i$ is the weight shown in Figure~\ref{fig:zbin} corresponding to the completeness correction with respect to the NVSS.

We present the luminosity function for two different redshift ranges. For the range 
0.005$<$z$<$0.3 (average $<z>$=0.07),  comprising 60 radio galaxies, and for the range 1$<$z$<$2.5 (with average $<z>$=1.53), containing 37 radio galaxies.
A luminosity function at intermediate redshifts is not presented due to the small number of detections. For either redshift range, we present both the luminosity function measured using detections only, and an alternative version using the combination of detections and upper-limits for non-detections. The H$_2$ luminosity (and consequently mass) function is shown in the left panel of Figure~\ref{fig:lco} for $z<$0.3 and in the right panel of the same figure for 1$<z<$2.5. 

For the local luminosity function, we have enough number statistics and  bins to fit the datapoints in logarithmic scale with an analytical   \citet{schechter76} function, as defined in \citet{riechers19}:
\begin{equation}
\log\Phi(L^\prime) = \log\Phi^* + \alpha \log\left( \frac{L^\prime}{L^*} \right) - \left( \frac{L^\prime}{\ln(10)L^*} \right)   + \log(\ln(10)).
\end{equation}
In this equation, $\Phi(L^\prime)$ is the number of galaxies per comoving volume per luminosity bin, $\Phi^*$ is the scale density of galaxies per
unit volume; $L^*$ is the scale luminosity defining the knee of
the LF and $\alpha$ is the slope of the faint end of the luminosity function.  The best Schechter function fit to our data is shown in Figure~\ref{fig:lco} as a dashed line, with 95\% confidence intervals shown as shaded areas. We found a negative slope of $\alpha$=-0.6, scale luminosity $L^*$ of 5.83$\times$10$^9$\,K\kms pc$^2$, and  turn over at $\Phi^*$=2.39$\times$10$^{-7}$\,Mpc$^{-3}$dex$^{-1}$.

For the local Universe, we compared our results with those from the eXtended CO Legacy Database for GASS (XCOLD GASS) survey \citep{saintonge17, fletcher21}, the CO survey of  FIR-selected galaxies from the IRAS bright galaxy sample \citep{keres03}, and the empirical CO luminosity function derived from the \textit{Herschel} IR luminosity \citep{vallini16}. Figure~\ref{fig:lco} shows that the gas mass content locked up in local radio galaxies is $\sim$2-4 orders magnitude less than that in normal star-forming galaxies, with the gap closing at high masses. The same applies when comparing the CO-ARC results to those of the CO observations in the blind ALMA Spectroscopic Survey of the \textit{Hubble} Ultra Deep Field (ASPECS LP) \citet{decarli19} (orange shaded boxes in Fig.~\ref{fig:lco}) for the average $z$ of each bin. ASPECS presents CO luminosity functions for galaxies of any kind, from the local Universe to $z\lesssim$4 and covers an area of 4.6\,arcmin$^2$. Deviations in the (bright-end) shape are seen for the sub-millimeter selected galaxies from the {\sc shark} semi-analytical model predictions by \cite{lagos20}. 

The comparison with the \textit{s}pectro-\textit{p}hotometric \textit{r}ealisations of \textit{i}nfrared-selected \textit{t}argets at all-\textit{z} (SPRITZ) simulation predictions \citep{spritz} turns out to be very instructive, as SPRITZ galaxies are separated in different populations including: unobscured and obscured AGN-dominated systems (hereafter AGN1 and AGN2, respectively) with little star formation, and composite systems of AGN and star-forming galaxies. One population of composite systems is similar to spirals hosting a low luminostity AGN (SF-AGN) and the other population resembles starburst galaxies hosting an obscured AGN (SB-AGN). 

The CO luminosity function in the local Universe interestingly shows that the number of radio galaxies with H2 gas detection is only little lower than that of pure type 1 or 2 AGN. The comparison between our findings and the SPRITZ simulation at low-$z$ shows that one on every $\sim$four RGs (i.e. a comparable number of RGs) have similar gas reservoirs to pure type 1 or type 2 AGNs. Surprisingly, even the number of RGs with large gas reservoirs at the bright-end of the COLF ($>$10$^{10}$M$_{\odot}$) is not considerably lower, so RGs are not entirely red and dead as it is often believed to be the case. The agreement between the molecular gas content in RGs and the AGN-dominated populations with little star formation (AGN1 and AGN2) from SPRITZ point out that we are comparing AGN populations where the star formation does not dominated the SED, since our $q_{\rm 24}$ selection criteria described in Section~\ref{subsec:sample_arc} assures that we selected galaxies whose radio emission is associated to black hole accretion instead of star formation. A deficit of gas at high masses is only seen for the very bright comparison sample, made to be representative of the 6C radio survey, flux limited to 3\,Jy (green points of Figure~\ref{fig:lco}, the bright sample properties are listed in Table~\ref{tab:bright} in the Appendix). At low CO luminosities, the SPRITZ simulations do not have predictions for $L^\prime<10^{7.25}$\,K\kms pc$^2$, although the shape of our Schechter function fit appears to agree with an extrapolated curve of the SPRITZ AGN1 and AGN2 predictions at lower luminosities.

\begin{figure*}
\centering
\includegraphics[width=\linewidth]{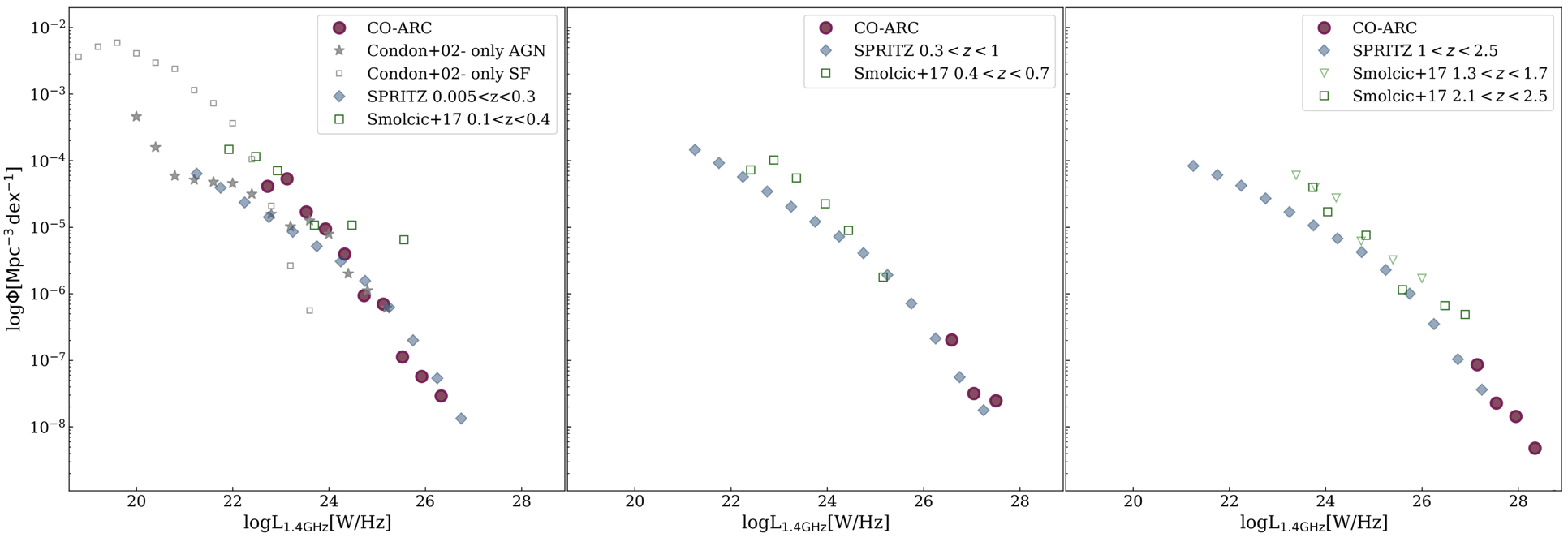}
\label{fig:l14}
\caption{Radio luminosity function at 1.4\,GHz for the low (left), intermediate (middle) and high (right) redshift bins. We also report the luminosity function of \citet{condon02} in the local Universe for star-forming and AGN population as open squares and stars, respectively. The SPRITZ predictions for radio-loud AGN are shown in gray diamonds \citet{spritz}, as well as the 1.4GHz luminosity function from the radio AGN survey in the VLA-COSMOS field \citep[][open green squares]{smolcic17} for each redshift bin.}
\label{fig:giga}
\end{figure*}

At high-$z$, the gas mass function of detected RGs with CO emission is lower (by up to 2.5 orders of magnitude) than that of local RGs with a corresponding mass content. It is even lower compared to other AGNs from the SPRITZ survey, and all galaxies in the ASPECS LP \citet{decarli19}. This discrepancy largely originates from the fact that only a small number of RGs from the 1.4\,GHz luminosity function are detectable in radio wavelengths at high-$z$. One of possible reason is due to the fact that the continuum spectrum of RGs at $\sim$GHz frequencies decreases with frequency and therefore it suffers from a strong K-correction, contrary to the CO lines  and continuum at mm wavelength, whose fluxes increase with frequency, leading to a negative K-correction and the possibility to be detected more easily at high z. Another explanation comes from our survey flux-limit of 0.4\,Jy, implying that only the most extremely luminous galaxies at the bright-end of the radio luminosity function can fit this criteria (see Figure~\ref{fig:giga} and text below). In other words, bright RGs are rarer to find, so our result needs to be roughly corrected with respect to the number density of undetectable RGs.

A rough attempt to correct this bias is made from the fraction of RGs in the inaccessible part of the 1.4\,GHz luminosity function, assuming a similar gas mass distribution throughout. For this reason, we created the 1.4 GHz luminosity function of each $z$ bin (Figure~\ref{fig:giga}) and identified what part of it our observations cover. At low-$z$, our survey covers a large part of the luminosity range of the SPRITZ simulations or of the NVSS and UGC galaxy samples \citet{condon98, condon02}. However, at high-$z$, the integral of the covered 1.4GHz luminosity function is much lower than that of the SPRITZ simulations, indicating that we only see the brightest of every $\sim$7000 galaxies. Therefore $\sim$7000 galaxies are missed, and based on our previous results, 35\% of them would, on average, contain detectable gas fractions. When accounting for the $\sim$2000 missing faint RGs in this way, we find that the gas mass content would be significantly higher than that in the local Universe, and comparable to that of other populations at high-z.

\begin{figure}
\resizebox{\hsize}{!}{\includegraphics{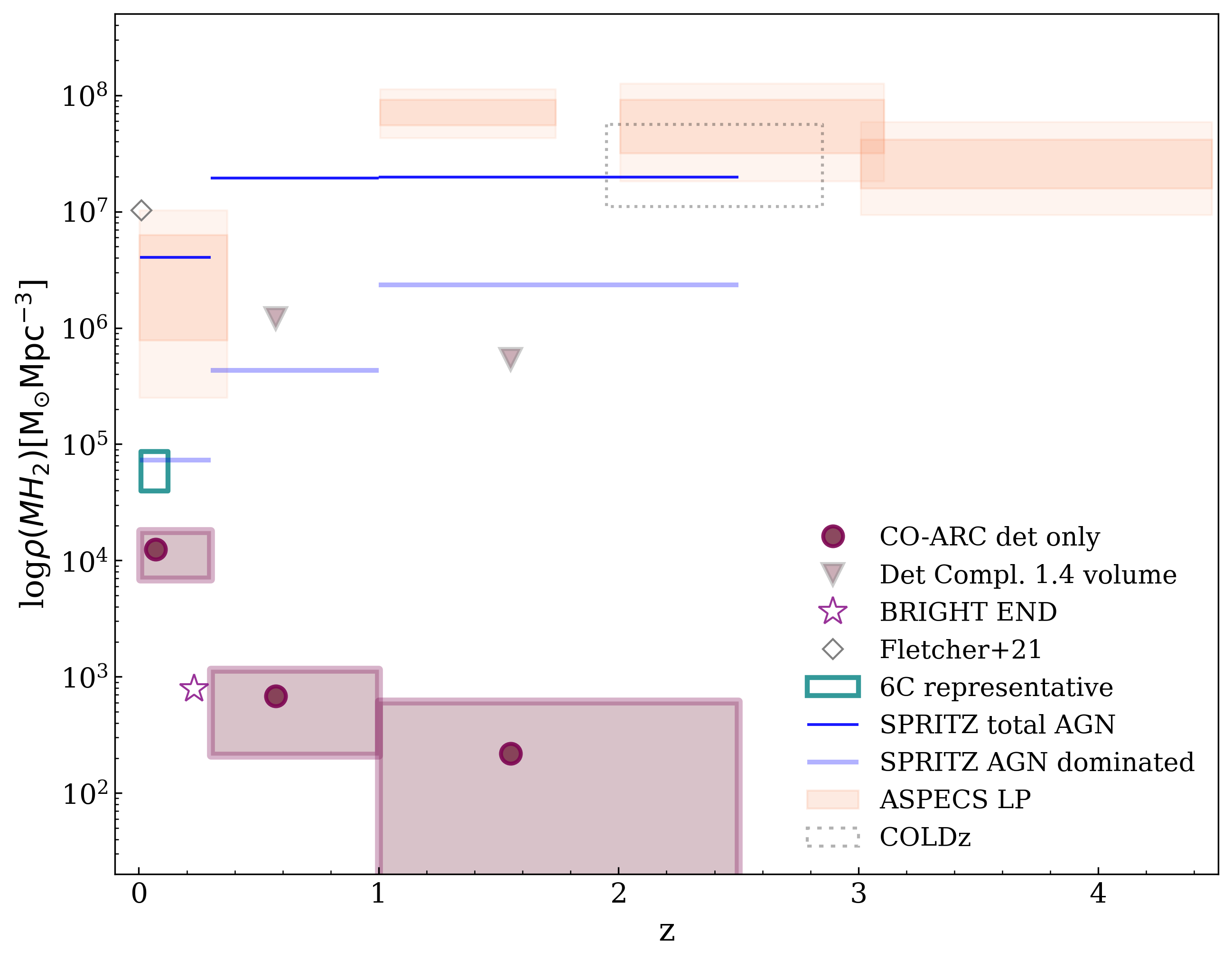}}
\caption{Evolution of the cosmic molecular gas density, $\rho(M\rm{_{H_2}})$, with redshift. Our results derived directly from the detection of the COLF are shown as the purple points and corresponding uncertainties boxes. The extrapolation of the number density of missing sources in the intermediate and high $z$ bins due to our flux limit is shown in the filled triangles. The $\rho(M\rm{_{H_2}})$ for the representative sample of the 6C catalog is indicated in the green box, and for the very bright end of our radio luminosity function ($L_{1.4GHz}>3\times10^{25}$\,W/Hz) is shown in the purple star. The light orange boxes indicate the results from the ASPECS LP survey \citep{decarli19} and the dotted box from the COLDz \citep{riechers19}. The dark and light blue lines represent the predictions from SPRITZ from all AGN hosts, including SF-AGN and SB-AGN, and the AGN-dominated (only AGN1+AGN2) populations.} 
\label{fig:rho}
\end{figure}

\section{Discussion: evolution of the M$_{\rm H_2}$ content of gas in radio galaxies with $z$}

By integrating $L'\Phi(L')$ over all luminosity function bins and converting it into mass, we obtain the molecular gas content per comoving  Mpc$^3$ of the Universe, $\rho(M\rm{_{H_2}})$, at different epochs (Figure~\ref{fig:rho}). Indeed, this plot shows that the derived molecular gas content of bright enough galaxies to be detected at 1.4 GHz (circles) shows a decrease  with $z$ due to the Malmquist bias. When correcting for the density of missing RGs as indicated above, i.e., by multiplying the derived high-$z$ $\rho(M\rm{_{H_2}})$ by $\sim$2000, then the corrected value is less than an order of magnitude below that of AGN1 plus AGN2, as in the local Universe. So, under the rough assumption that detectable and undetectable RGs have a similar gas mass distribution, the amount of gas locked-up in RGs and AGNs is similar at $z>1$ - as the local Universe. Likewise, in the intermediate $z$, we only see the brightest of every 5000 galaxies, from which we deduce the appropriate $\rho(M\rm{_{H_2}})$ correction.  However, we draw no further conclusions from this bin, as it suffers from small number statistics (only 2 detections out of 23 sources there). 

We further investigated our findings by only examining the most luminous objects at different epochs. Due to the $z$ evolution of luminosities \citep{pracy16}, we avoided directly comparing equal-L galaxies at different $z$. Instead we compared the gas mass content of the brightest of every few thousand RGs at any given $z$. For comparison, we cut the local 1.4GHz luminosity function at 9$\times$10$^{25}$  
W/Hz, thus using only the brightest of every 7000 
local galaxies. We then reconstructed the CO luminosity function and local Universe gas mass density using these galaxies. The outcome of this computation is the open star in Figure~\ref{fig:rho}. We find that the same amount of molecular gas is locked up in the brightest 1/5000-1/7000 radio galaxies at any given redshift. This result indicates that the net effect of intergalactic medium inflows on one hand and black hole feedback and star formation on the other hand, integrated over time, leads to the same amount of residual molecular gas in the brightest radio galaxies. It is, thus, plausible that radio-mode feedback indeed acts as a ``maintenance'' mode. To further check this, an independent measurement of the star formation rate in these galaxies is needed. This will be presented in a future paper from the fitting of the IR data.

\section{Summary}
\label{sec:summary}

To quantify the gas mass content of radio galaxies at different epochs, we examined a sample of 120 radio galaxies from the local Universe up to z$\sim$2.5. Our sample was meticulously constructed to be complete, representative of the 1.4\,GHz NVSS survey when flux limited to 0.4\,Jy. All galaxies in it have radio emission associated with black hole activity, as indicated by a radio-to-infrared ratio q$_{\rm 24\mu m}<$0.5. Of the sources, 66 were ALMA calibrators, originating from the ALMA Radio-Source Catalogue, and were often observed as such. Another 54 sources, often comprising more extended radio emission,
were compiled from the literature. For all galaxies, CO (1-0) up to (4-3) data were used for the gas mass content derivation. Our main findings can be summarized as follows.

\begin{itemize}
\item

We detected CO emission in 35\% of the sample galaxies, with molecular masses ranging from $\rm 1\times 10^7 <M_{H_2}< 7 \times 10^{10} M_{\odot}$ at low-$z$ and $\rm 2\times 10^{10} <M_{H_2}< 7 \times 10^{11} M_{\odot}$ at high-$z$. Two sources with outflows were seen among the 17 sources with newly reduced ALMA data.
\item 
One quarter of the 1$<$$z$$<$2.5 radio galaxies contain at least as much molecular gas as a typical galaxy halo of 10$^{13}$-10$^{14}$ $M_{\odot}$. Therefore, a large number of radio galaxies at high-$z$ would reside at a "normal" or even starbursty for that epoch host. Still, at any $z$, most radio galaxies are more depleted and thus evolved than the typical galaxy of a simulated halo. 
\item
The CO luminosity function in the local Universe interestingly shows that the number of radio galaxies with \htwo\ gas detection is only little lower than that of pure type 1 or 2 AGN.
\item
At 1$<$$z$$<$2.5, the CO luminosity function is significantly lower than that predicted by simulations for other galaxy types, because only hard-to-find bright radio galaxies are detectable. A rough correction of the fraction of missing sources of the 1.4 GHz luminosity function implies that the CO luminosity function would be 2000 times higher. This would close the gap between the various high-$z$ populations and it would imply that the gas mass locked up in high-$z$ galaxies is considerably higher than in the local Universe
\item
The gas mass content of the brightest 1/5000-1/7000 radio galaxies at any $z$ is similar. It is plausible that these molecular gas reservoirs are maintained at similar levels thanks to both the consumption of gas due to star formation and the delay of inflows from the intergalactic medium due to radio activity.
\end{itemize}
This work demonstrate a potential case of exploring the wealth of the ALMA archival observations, including calibrators, for scientific purposes. The CO luminosity function of radio galaxies derived here might be useful for benchmarking of cosmological simulations and constraining the gaseous content of radio galaxies at the local Universe up to the star formation/activity history peak at $z\lesssim$2.5, where feedback could be more crucial for galaxy evolution.

\begin{acknowledgements}

We thank the referee for the constructive review and comments. KMD, AA, MP, JAFO, and IR acknowledge financial support by the Hellenic Foundation for Research and Innovation (HFRI), under the first call for the creation of research groups by postdoctoral researchers that was launched by the General Secretariat For Research and Technology under project number 1882 (PI Dasyra, Title: "Do massive winds induced by black-hole jets alter galaxy evolution? Evidence from galaxies in the ALMA Radio-source Catalog (ARC)"). AA also acknowledges the projects ``Quantifying the impact of quasar feedback on galaxy evolution'', with reference EUR2020-112266, funded by MICINN-AEI/10.13039/501100011033 and the European Union NextGenerationEU/PRTR, and from the Consejer\' ia de Econom\' ia, Conocimiento y Empleo del Gobierno de Canarias and the European Regional Development Fund (ERDF) under grant ``Quasar feedback and molecular gas reservoirs'', with reference ProID2020010105, ACCISI/FEDER, UE.
JAFO acknowledges the financial support from the Spanish Ministry of Science and Innovation and the European Union -- NextGenerationEU through the Recovery and Resilience Facility project ICTS-MRR-2021-03-CEFCA.  IR also acknowledges support from the UK Science and Technology Facilities Council through grant ST/S00033X/1. This paper makes use of the ALMA data with project IDs listed in Tables~\ref{tab:proj} and 1. ALMA is a partnership of ESO (representing its member states), NSF (USA) and NINS (Japan), together with NRC (Canada) and NSC and ASIAA (Taiwan), in cooperation with the Republic of Chile. The Joint ALMA Observatory is operated by ESO, AUI/NRAO and NAOJ. The National Radio Astronomy Observatory is a facility of the National Science Foundation operated under cooperative agreement by Associated Universities, Inc. We made use of the NASA/IPAC Extragalactic Database (NED), and of the VizieR catalogue access tool, CDS, Strasbourg, France. This research made use of Astropy, a community developed core Python package for Astronomy.

\end{acknowledgements}


\bibliographystyle{aa} 
\bibliography{CO-ARC.bib}

\begin{appendix} 

\section{Moment maps and spectra}

\begin{figure*}[b]
\centering
\includegraphics[width=17cm]{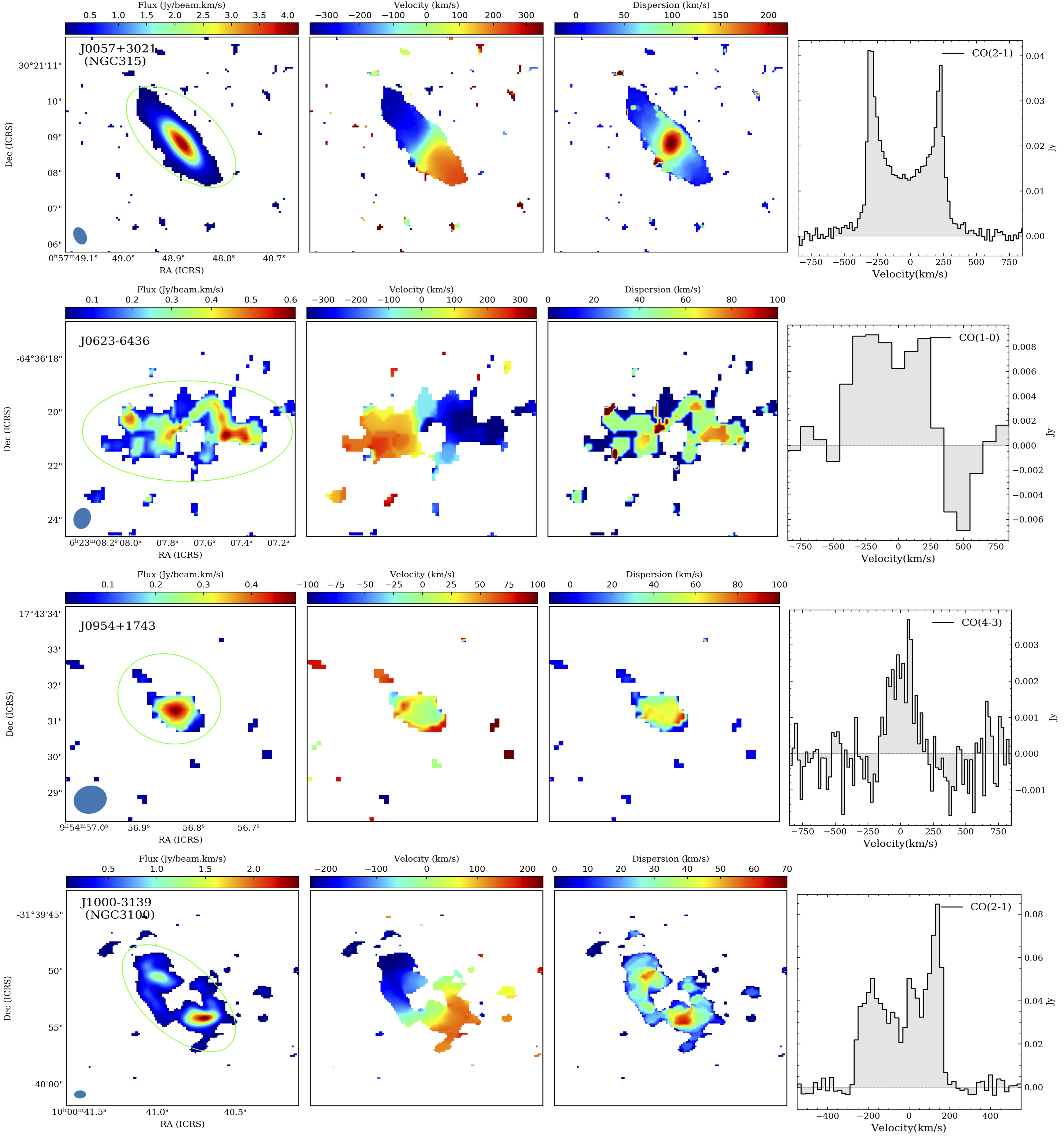}
\caption{Integrated intensity (moment 0; left panels), mean line-of-sight velocity (moment 1; middle-left panels), velocity dispersion (moment 2; middle-right) maps and spectra (right panels) of the 17 detections in the CO-ARC sample. The synthesized beam is shown as a blue ellipse in the bottom-left corner of each moment 0 map. The bar to the top of each moment map shows the color scales (in Jy~beam$^{-1}$~km~s$^{-1}$ and km~s$^{-1}$ for moment 0 and moment 1/2 maps, respectively). East is to the left and north to the top. The observed CO transition is indicated in the top-right corner of each spectral profile. The aperture within which the illustrated spectra have been extracted are overlain in green in each moment 0 maps.}
\label{fig:detall}
\end{figure*}

We present in Figure~\ref{fig:detall} the integrated intensity, mean line-of-sight velocity and velocity dispersion maps of the 17 detections out of the 66 CO-ARC sources listed in Table~\ref{tab:detections} and discussed in Section~\ref{mommaps}.

\begin{figure*}[h!]
\centering
\includegraphics[width=17cm]{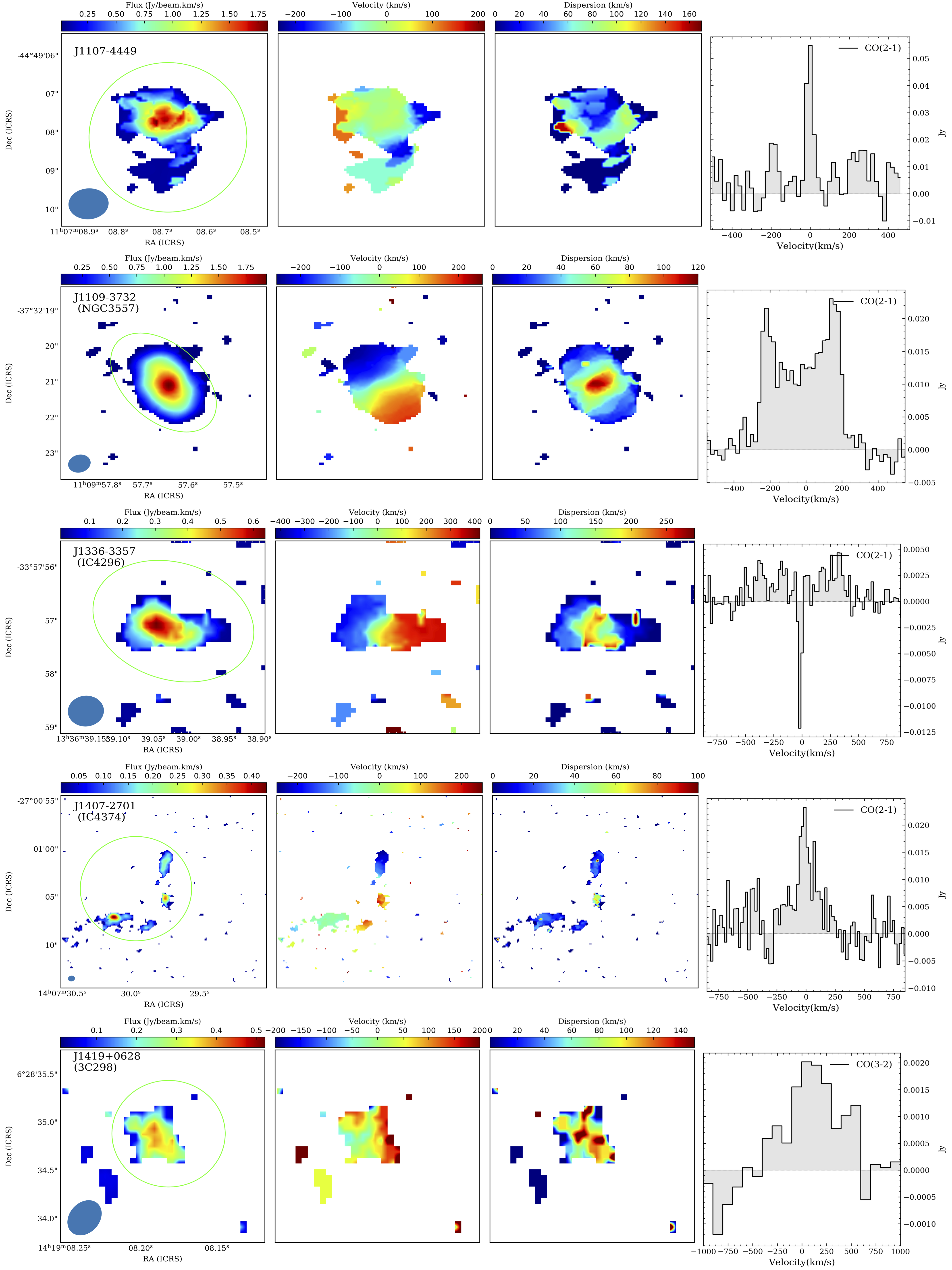}
{ \rm \bf Fig.~\ref{fig:detall}\,-\,continued. }
\end{figure*}

\begin{figure*}[h!]
\centering
\includegraphics[width=17cm]{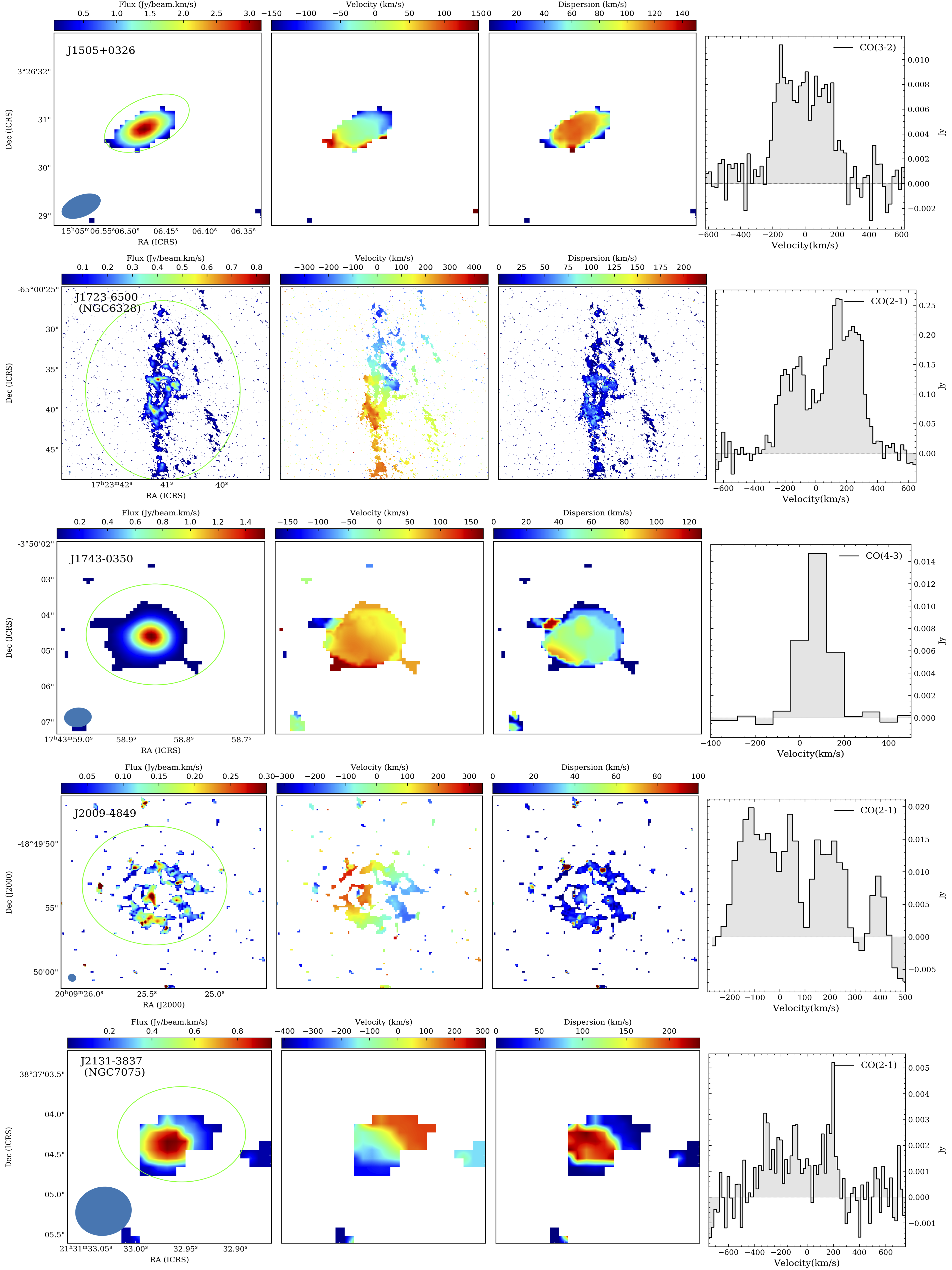}
{ \rm \bf Fig.~\ref{fig:detall}\,-\,continued. }

\end{figure*}

\begin{figure*}[h!]
\centering
\includegraphics[width=17cm]{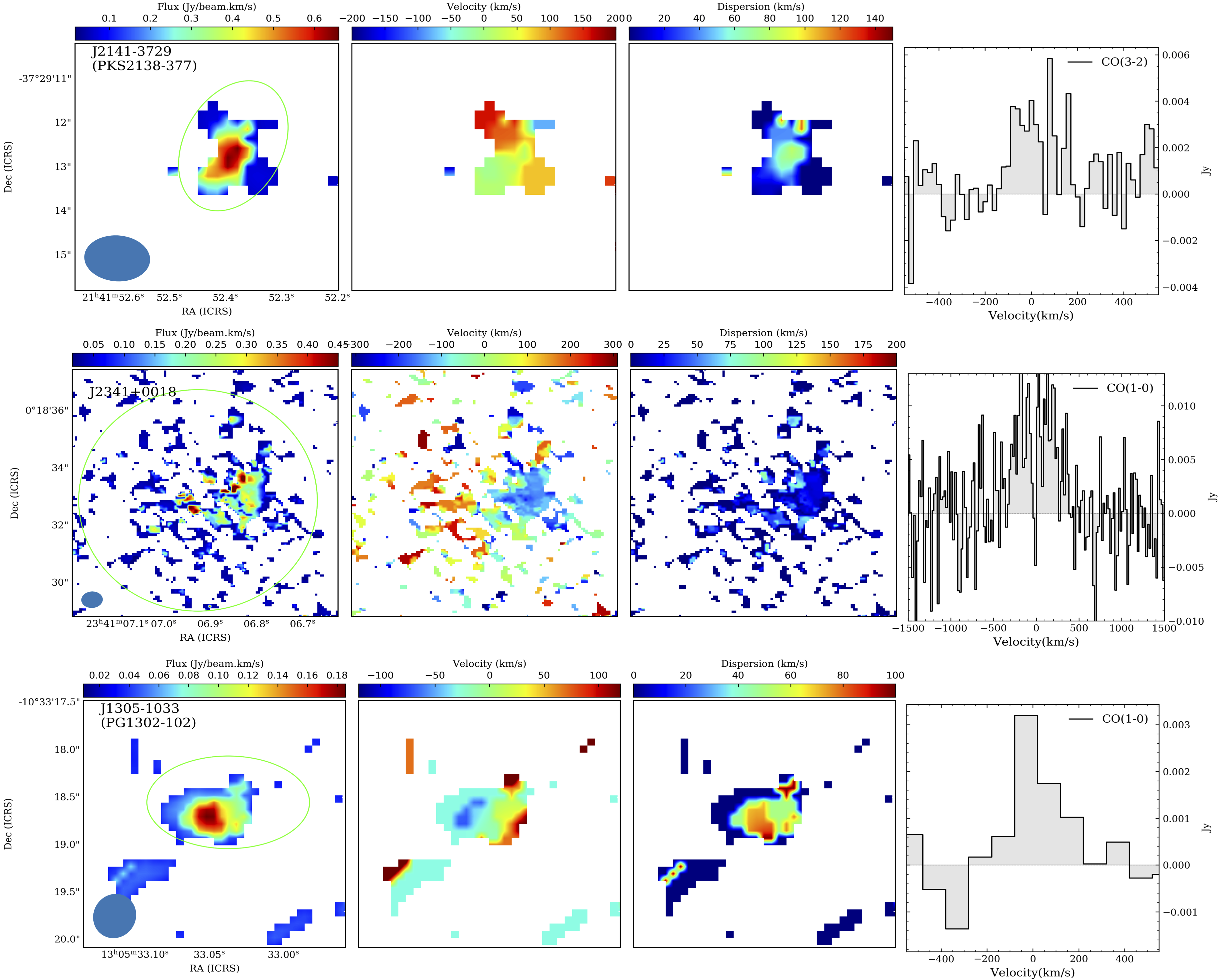}
{ \rm \bf Fig.~\ref{fig:detall}\,-\,continued. }
\end{figure*}

\clearpage
\newpage

\section{Summary of the ALMA data}

In Table~\ref{tab:proj} we report main properties (name, coordinates, redshift and 1,4\,GHz fluxes) of the the 66 ALMA CO-ARC sources and the respective ALMA projects, integration times and scan intents used in this work.

\onecolumn
\begin{landscape}
\fontsize{8}{10}\selectfont 
\setlength\tabcolsep{1pt}
\setlength\LTcapwidth{\textwidth} 
\setlength\LTleft{0pt}            
\setlength\LTright{0pt} 
\begin{longtable}{@{} lcccccccccccc @{}}
\caption{CO-ARC sample ancillary information and information on the reduced ALMA observations.}\\
\hline \hline
     Galaxy &  Coordinates & z & f$_{1.4}$ &  PID & intent & PID & intent &  PID & intent & PID & intent & redshift  \\
    &  &  &  &  \cooz\ & \cooz\ & \coto\ & \coto\ & \cott\ & \cott\ & \coft\ & \coft\ & reference  \\
    &  (J2000) &  & (Jy) & &  &  & & & & &   \\
\hline 
\endhead
\hline
\endfoot  
J1000-3139,NGC3100 &   10:00:40.84 -31:39:52.35 & 0.009 & 0.558 & - &   - & 2015.1.01572.S &  28.72$^T$ &  \makecell[t]{-} &  - & - & - &  \citet{Jones2009} \\
 J1109-3732,NGC3557 &  11:09:57.64 -37:32:21.02 & 0.010 & 0.401 & - &   - &  \makecell[t]{2015.1.01572.S} &  \makecell[t]{22.68$^T$} &- & - & - & - &  \citet{Jones2009} \\
 J1336-3357,IC4296 &  13:36:39.03 -33:57:57.30 & 0.012 & 3.131 & - & -  & \makecell[t]{2015.1.01572.S} &  \makecell[t]{ 51.4$^T$ } &- & - & - & - &  \citet{Smith2000} \\
J1723-6500,NGC6328 &  17:23:41.03 -65:00:36.48 & 0.014 & 3.636 & - & -  & 2015.1.01359.S & 113$^T$ &- & - & - & - & \citet{Fosbury1977} \\
J1945-5520,NGC6812 &  19:45:24.22 -55:20:48.84 & 0.015 & 0.690 & 2011.0.00046.S & 5.04$^P$ &- & - &- & - & - & - &  \citet{Jones2009} \\
J0057+3021,NGC0315 &  00:57:48.88 +30:21:08.81 & 0.017 & 1.550 & - & -  & \makecell[t]{2017.1.00301.S} &  84.7$^T$ &- & - & - & - & \citet{Peterson1979} \\
J1301-3226,ESO443-G024 &   13:01:00:79 -32:26:28.97 & 0.017 & 1.298 & - & -  &2015.1.01572.S &  24.2$^T$ &- & - & - & - &  \citet{Smith2000} \\
 J2131-3837,NGC7075 &  21:31:32:98 -38:37:04.65 & 0.018 & 0.668 & - & -  & 2015.1.01572.S &  24.7$^T$ &- & - & - & - &  \citet{Jones2009}\\
 J1407-2701,IC4374 &   14:07:29.76 -27:01:04.39 & 0.022 & 0.706 & \textcolor{gray}{2017.1.00629.S} &   28.7$^T$ & 2015.1.00644.S &  87.7$^T$ &- & - & - & - &  \citet{Smith2000} \\
J1516+0701,UGC09799  &  15:16:44.49 +07:01:17.83 & 0.035 & 3.934 & - & -  &2015.1.00627.S &  81.6$^T$ & - & - & - & - &  SDSS dr13 \citet{Albareti2017} \\ 
 J2009-4849,HB89 2005-489  &  20:09:25.39 -48:49:53.72 & 0.071 & 1.117 & - & - & 2013.1.00523.S &  12.1$^P$ &- & - & - & - & \citet{Richter2016} \\
J1008+0029,PKS1005+007  &  10:08:11.44 +00:29:59.60 & 0.098 & 0.600 & 2019.1.00102.S &  14.1$^P$  &- & - &- & - & - & - & \citet{Owen1995} \\
J1221+2813,W Com &  12:21:31.69 +28:13:58.50 & 0.103 & 0.758 &\makecell[t]{2015.1.00727.S} &   90.7$^T$ &- & - &- & - & - & - & \citet{Shaw2013} \\
   J0623-6436,WISEJ062307-643620 &  06:23:07.71 -64:36:20.6 & 0.129 & 0.400 & 2015.1.01522.S & 5.24$^T$ &- & - &- & - & - & - & \citet{Pietsch1998} \\
J1217+3007,FBQSJ1217+3007 &  12:17:52.08 +30:07:00.64 & 0.130 & 0.539 &\makecell[t]{2015.1.00727.S \\ 2015.1.00820.S} &  \makecell[t]{ 9.6$^P$\\ 3.0$^P$} &- & - &- & - & - & - & \citet{Landoni2020} \\
 J1427+2348,FBQSJ1427+2348 &   14:27:00.39 +23:48:00.04 & 0.160 & 0.452 &\makecell[t]{2017.1.00616.S} &   18.14$^P$ &- & - &- & - & - & - & \citet{Richter2016} \\
 J1332+0200,3C287.1 &  13:32:53.27 +02:00:45.70 & 0.216 & 3.000 &\makecell[t]{2017.1.01359.S \\ 2015.1.00804.S} &  \makecell[t]{ 12.7$^P$ \\ 3.53$^P$} &- & - &- & - & - & - & SDSS dr13 \citet{Albareti2017} \\
 J1356-3421,PKS1353-341 & 13:56:05.39 -34:21:10.84 & 0.223 & 0.811 & 2017.1.00629.S &   20.7$^T$ &- & - &- & - & - & - & \citet{Veron2000} \\
J0943-0819,PKS0941-08 &   09:43:36.94 -08:19:30.81 & 0.228 & 1.828 & 2012.1.00915.S &   10.6$^P$ &- & - &- & - & - & - & \citet{Son2012} \\
J1220+0203,UM\,492 &  12:20:11.88 2:03:43.2 & 0.240 & 0.548 & \makecell[t]{2016.1.00140.S \\ 2016.1.01481.S } & \makecell[t]{2.1$^P$ \\ 0.6$^C$} &- & - & \textcolor{gray}{2016.1.00994.S} & 1.5$^P$ & - & - & \citet{Hewett2010} \\
J1547+2052,PG,1545+210 &  15:47:43.54 +20:52:16.61 & 0.264 & 1.456 & 2017.1.01249.S & 4.5$^C$ &- & - &- & - & - & - &   SDSS dr6 \citet{Adelman-McCarthy2008} \\
 J2341+0018,PKS2338+000 & 23:41:06.91 +00:18:33.34 & 0.277 & 0.441 & 2017.1.00629.S &   21.67$^T$ &- & - &- & - & - & - & \citet{Katgert1998} \\
J1305-1033,PG\,1302-102 &  13:05:33.01 -10:33:19.38 & 0.278 & 0.739 & 2015.1.00329.S &   38.3$^T$ &- & - &- & - & - & - & \citet{Wisotzki2000} \\
 J0242-2132,PKS0240-217 & 02:42:35.91 -21:32:25.94 & 0.314 & 1.003 & 2017.1.00629.S &   21.7$^T$ &- & - &- & - & - & - & \citet{Wright1983} \\
 J0006-0623,PKS0003-066 & 00:06:13.89 -06:23:35.34 & 0.347 & 1.704 &  \makecell[t]{2012.1.00080.S \\ 2017.1.00629.S } & \makecell[t]{6$^B$ \\ 5$^P$ }& - & - & - & - & - & - &  \citet{Jones2009} \\
  J1505+0326,HB89 1502+036 & 15:05:06.48 +03:26:30.81 & 0.408 & 0.416 & - & -  &- & - & \makecell[t]{2015.1.00971.S} &  7.06$^P$ & - & - &  \citet{Hewett2010} \\
 J0748+2400,PKS0745+241 & 07:48:36.11 +24:00:24.11 & 0.410 & 1.035 & - & -  &- & - & 2015.1.01178.S &   8.06$^P$ & - & - &  \citet{Hewett2010} \\
J0510+1800,PKS0507+17 &   05:10:02.37 +18:00:41.58 & 0.416 & 0.711 & - & -  &- & - & \makecell[t]{2012.1.00275.S \\ 2012.1.00193.S } & \makecell[t]{19.7$^{P}$ \\ 7.6$^P$} & - & - & \citet{Perlman1998} \\
J2349+0534,WISEJ234921+053439 &  23:49:21.05 +05:34:39.87 & 0.419 & 0.445 & - & -  &- & - &2019.1.01229.S &  12.1$^P$ & - & - & \citet{Landoni2020} \\
 J2141-3729,PKS2138-377 &   21:41:52.45 -37:29:12.99 & 0.423 & 0.480 & - & -  &- & - & 2017.1.00255.S &   3.02$^P$ & - & - &  \citet{Jones2009} \\
J0914+0245,PKS0912+029 &  09:14:37.91 +02:45:59.25 & 0.427 & 0.540 & - & -  &- & - & 2011.0.00016.S &  36.89$^P$ & - & - &  \citet{Hewett2010} \\
  J1038+0512,PKS1036+054 &  10:38:46.78 +05:12:29.09 & 0.473 & 0.601 & - & -  &2017.1.01558.S &   3.02$^P$ &- & - & - & - & \citet{Healey2008} \\
J0940+2603,B2 0937+26 &  09:40:14.72 +26:03:29.95 & 0.498 & 0.460 & - & -  &- & - &  \makecell[t]{2017.1.00277.S } &   6.05$^P$ & - & - & \citet{Glikman2007} \\
   J1610-3958,WISEJ161021-395858 &  16:10:21.88 -39:58:58.33 & 0.518 & 0.598 & - & -  &- & - & \makecell[t]{2015.1.00791.S \\ 2017.1.00704.S} &   \makecell[t]{ 9.07$^P$ \\ 6.05$^P$} & - & - &  \citet{Landt2001} \\
J2239-5701,PKS2236-572 & 22:39:12.08 -57:01:01 & 0.569 & 0.432 & - & -  &- & - & 2016.2.00115.S &  18.13$^P$ & - & - &  \citet{Titov2011} \\
J1058-8003,PKS1057-79 &  10:58:43.31 -80:03:54.16 & 0.581 & 0.997 & - & -  & - & - & 2017.1.00886.L &   60$^P$ & - & - &  \citet{Sbarufatti2009} \\
J0106-4034,HB89 0104-408 & 01:06:45.11 -40:34:19.96 & 0.584 & 0.768 & - & -  &- & - & - & - & 2016.1.00133.T &   8.06$^P$ &  \citet{White1988} \\
 J0217-0820,PKS0214-085 &  02:17:02.66 -08:20:52.35 & 0.607 & 0.459 & - & -  & 2018.1.00478.S &   4.02$^P$ &  \textcolor{gray}{2017.1.00562.S} &  7.06$^P$ & - & - &  \citet{Hewett2010} \\
  J2320+0513,HB89 2318+049 &   23:20:44.86 +05:13:49.95 & 0.622 & 0.735 & - & -  &2018.1.01533.S &   7.56$^P$ &- & - & - & - &  \citet{Schmidt1977} \\
J2000-1748,HB89 1958-179 &  20:00:57.09 -17:48:57.67 & 0.652 & 0.751 & - & -  &  \makecell[t]{2018.1.00576.S } &  5.04$^B$ & - &  - & - & - &  \citet{Drinkwater1997} \\ 
 J1010-0200,4C\,-01.21  &  10:10:52.59 -02:00:19.57 & 0.890 & 0.680 & - & -  &- & - &- & - &\makecell[t]{ 2018.1.00583.S} &  \makecell[t]{3.53$^P$} &  \citet{Croom2004} \\
J0329-2357,PKS0327-241 &  03:29:54.08 -23:57:08.77 & 0.895 & 0.683 & - & -  &- & - &- & - &2017.1.00685.S &  22.7$^P$ &  \citet{Baker1999} \\
J0946+1017,WISEJ094635+101706 &   09:46:35.07 +10:17:06.13 & 1.004 & 0.406 & - & -  &2013.1.00432.S &   3.02$^P$ &- & - & - & - &  \citet{Hewett2010} \\ 
 J0909+0121,HB89\,0906+015 &  09:09:10.09 +01:21:35.62 & 1.025 & 0.954 & - & -  &2017.1.00719.S &   3.53$^P$ &- & - &  \textcolor{gray}{2015.1.01455.S} & \textcolor{gray}{6.05$^P$} &  \citet{Hewett2010} \\
 J1351-2912,PKS\,1348-289 &  13:51:46.84 -29:12:17.65 & 1.034 & 0.523 & - & -  & \makecell[t]{2015.1.00645.S } &  \makecell[t]{24.2$^P$} &- & - & \textcolor{gray}{2016.1.00164.S} & \textcolor{gray}{36.3$^P$} & \citet{Hook2003} \\
 J1743-0350,HB89\,1741-038 & 17:43:58.86 -03:50:04.56 & 1.054 & 1.652 & - & -  &  - & - &- & - & \makecell[t]{2015.1.00492.S} & \makecell[t]{117$^{Pol}$ } &  \citet{Zafar2013} \\
J0125-0005,UM\,321 &   01:25:28.84 -00:05:55.93 & 1.076 & 1.275 & - & -  & 2015.1.00570.S &  16.1$^P$ &- & - & - & - &  \citet{Hewett2010} \\
J0239-0234,HB89 0237-027 &   02:39:45.47 -02:34:40.91 & 1.116 & 0.400 & - & -  &  - &  - &- & - & 2016.1.00232.S &  33.26$^P$ & \citet{Fricke1983} \\
 J0837+2454,FBQSJ083740+245423 &  08:37:40.25 +24:54:23.12 & 1.126 & 0.514 & - & -  & \makecell[t]{2015.1.00804.S \\ 2016.1.00406.S } &  \makecell[t]{6.05$^P$ \\ 5.14$^C$ } &- & - & - & - &  \citet{Hewett2010} \\
 J0118-2141,HB89\,0116-219 &  01:18:57.26 -21:41:30.14 & 1.161 & 0.498 & - & -  &- & - &- & - & \makecell[t]{2015.1.00902.S} &  4.03$^P$ &  \citet{Hewitt1989} \\
 J0112-6634,PKS\,0110-668 &  01:12:18.91 -66:34:45.19 & 1.189 & 0.439 & - & -  &  \makecell[t]{2017.1.00280.S} &   4.03$^P$ &- & - & - & - &  \citet{Titov2011} \\
  J1304-0346,HB89\,1302-034 &  13:04:43.64 -03:46:02.55 & 1.250 & 0.958 & - & -  &\makecell[t]{2016.1.00922.S} &  9.07$^P$ &- & - & - & - & \citet{Wills1978} \\
J2134-0153,HB89\,2131-021 & 21:34:10.31 -01:53:17.24 & 1.285 & 2.040 & - & -  & 2019.1.00790.S &   9.1$^P$ & - & - & - & - &  \citet{Drinkwater1997} \\
 J1359+0159,HB89\,1356+022 &   13:59:27.15 +01:59:54.56 & 1.326 & 0.777 & - & -  & \makecell[t]{2016.1.00991.S} &  \makecell[t]{ 14.11$^P$} & - & - & - & - &  \citet{Hewett2010} \\
 J0529-7245,PKS\,0530-727 &  05:29:30.04 -72:45:28.51 & 1.340 & 0.400 & - & -  & \makecell[t]{2015.1.01388.S \\  2017.1.00093.S} & \makecell[t]{12.6$^P$ \\ 6.05$^P$ } &- & - & - & - & \citet{Gattano2018} \\
J1147-0724,PKS\,1145-071 &  11:47:51.55 -07:24:41.14 & 1.342 & 0.878 & - & -  & - & - &- & - &2018.1.00699.S &   8.05$^C$ & \citet{Wilkes1986} \\
  J0343-2530,PKS\,0341-256 &  03:43:19.52 -25:30:17.41 & 1.419 & 0.602 & - & -  &2016.1.00754.S &   6.05$^P$ &- & - & - & - & \citet{Hook2003} \\
 J1419+0628,3C\,298  & 14:19:08.18 +06:28:34.76 & 1.438 & 6.060 & - & -  &- & - & \makecell[t]{2013.1.01359.S \\ 2015.1.01090.S} &  \makecell[t]{ 27.2$^T$ \\ 24.2$^T$} & - & - &  \citet{Hewett2010} \\
J0954+1743,HB89\,0952+179 &  09:54:56.82 +17:43:31.22 & 1.477 & 1.184 & - & -  &- & - &- & - & 2017.1.01559.S & 151.2$^T$ &  \citet{Hewett2010} \\
J2056-4714,HB89\,2052-474 & 20:56:16.36 -47:14:47.63 & 1.489 & 2.949 & - & -  & \makecell[t]{ 2016.1.01481.S \\ 2017.1.00273.S \\ 2017.1.01100.S } & \makecell[t]{6.05$^B$ \\ 15.12$^B$  \\ 10.08$^B$ } & - & - & - & - & \citet{Jauncey1984} \\
J1520+2016,3C\,318 &  15:20:05.45 +20:16:05.80 & 1.572 & 2.427 & - & -  &- & - & 2017.1.01527.S &  69.6$^T$ & - & - &  SDSS dr13 \citet{Albareti2017} \\
 J1107-4449,HB89\,1104-445 &  11:07:08.69 -44:49:07.62 & 1.598 & 2.116 & - & -  & \makecell[t]{ 2012.1.00394.S \\ 2017.1.01678.S \\ 2017.1.00555.S } & \makecell[t]{33$^B$ \\ 5$^B$ \\ 5$^B$} & 2017.1.01202.S &  25.2$^B$ & - & - & \citet{PetersonB1979} \\
 J0219+0120,PKS\,0216+011 &  02:19:07.02 +01:20:59.87 & 1.623 & 0.518 & - & -  &  \makecell[t]{2013.1.00055.S \\ 2013.1.00279.S} &  \makecell[t]{6.05$^P$ \\ 6.05$^P$} &- & - & - & - & \citet{Allington-Smith1991} \\
  J1136-0330,4C\,-03.44 &   11:36:24.58 -03:30:29.50 & 1.648 & 0.400 & - & -  &2017.1.01694.S &   4.02$^P$ &- & - & - & - & \citet{Hook2003} \\
 J1146-2447,HB89\,1143-245 &  11:46:08.10 -24:47:31.2 &  1.94 &  0.885 & -& - & - & - & -& - & 2015.1.00851.S & 10.6$^P$ & \citet{Drinkwater1997} \\
 J0403+2600,HB89\,0400+258 & 04:03:05.59 +26:00:01.50 & 2.109 & 1.367 & - & - & - & - & - & - & 2018.1.00536.S  & 58.3$^P$ & \citet{Healey2008}  \\  
J0106-2718,HB89\,0104-275 & 01:06:26.08 -27:18:10.8 & 2.486 & 0.400 & -& - & - & - & \makecell[t]{2015.1.01487.S} & \makecell[t]{10.6$^P$} & - & - & \citet{Croom2004} \\
           \label{tab:proj}
\end{longtable}

\tablefoot{Columns are, from left to right: galaxy name, right ascension and declination coordinates,  redshift, 1.4\,Ghz flux, ALMA project identification code (PID) and integration time (in minutes) with the scan intents superscript - T for targets, P for phase calibrators, B for bandpass calibrators, C for check water-vapor calibrators and Pol for polarization calibrators - for the CO(1-0), CO(2-1), CO(3-2), and CO(4-3) transitions, and the reference for the spectroscopic redshift.}
\end{landscape}

\twocolumn
\section{Properties of the molecular gas in the literature data}

We present in Table~\ref{tab:biblio} the properties of the molecular gas measurements for the 54 literature sources that are part of our CO-ARC sample.
\begin{table*}[b]
\footnotesize
\centering
\caption{CO properties of 54 literature-based sources.}
\begin{tabular}{llllll}
\hline

   Galaxy  &    $z$ &  log(M$_{H_2}$) & log(L'$\rm_{CO(1-0)}$) & f$_{1.4 GHz}$  & reference \\
   & & (M$\rm_\odot$) & (K\kms pc$^2$) & (Jy) & \\
\hline
              NGC1399 &  0.005 &     7.47 &       6.83 &  0.639 &      Prandoni et al. 2010 \\
             IC1459 &  0.006 &  $<$6.00 &    $<$5.36 &  0.840 &         Ruffa et al. 2019 \\
            NGC4696 &  0.010 &     7.95 &       7.31 &  3.922 &      Olivares et al. 2019 \\
             NGC3801 &  0.011 &     8.74 &       8.10 &  1.141 & Ocaña-Flaquer et al. 2010 \\
             NGC7626 &  0.011 &  $<$7.90 &    $<$7.26 &  0.627 &          Leon et al. 2002 \\
             NGC 193 &  0.014 &  $<$8.08 &    $<$7.44 &  1.375 &          Leon et al. 2002 \\
          J0048+3157 &  0.015 &  $<$8.73 &    $<$8.09 &  0.401 &      Taniguchi et al 1990 \\
             NGC5127 &  0.015 &     7.90 &       7.26 &  0.500 & Ocaña-Flaquer et al. 2010 \\
             NGC5490 &  0.016 &  $<$8.73 &    $<$8.09 &  0.971 & Ocaña-Flaquer et al. 2010 \\
             NGC5141 &  0.018 &  $<$8.70 &    $<$8.06 &  0.638 & Ocaña-Flaquer et al. 2010 \\
              NGC541 &  0.018 &     8.40 &       7.76 &  0.826 & Ocaña-Flaquer et al. 2010 \\
             NGC741 &  0.019 &  $<$8.74 &    $<$8.10 &  0.635 &         Evans et al. 2005 \\
            NGC2329 &  0.019 &  $<$8.15 &    $<$7.51 &  0.512 &          Leon et al. 2002 \\
              IC1531 &  0.026 &     8.04 &       7.40 &  0.505 &         Ruffa et al. 2019 \\
               3C442 &  0.026 &     7.83 &       7.19 &  2.216 & Ocaña-Flaquer et al. 2010 \\
          PKS0718-34 &  0.028 &  $<$6.83 &    $<$6.19 &  1.909 &         Ruffa et al. 2019 \\
               3C465 &  0.030 &  $<$8.89 &    $<$8.25 &  6.134 & Ocaña-Flaquer et al. 2010 \\
      ESO422-G028 &  0.038 & $<$10.05 &    $<$9.41 &  1.500 &     Saripalli et al. 2007 \\
             Arp187 &  0.040 &    10.07 &       9.43 &  1.461 &         Evans et al. 2005 \\
          J0758+3747 &  0.041 &  $<$9.30 &    $<$8.66 &  1.734 & Ocaña-Flaquer et al. 2010 \\
               3C305 &  0.042 &     9.31 &       8.67 &  2.950 & Ocaña-Flaquer et al. 2010 \\
              NGC326 &  0.048 &     8.75 &       8.11 &  1.231 & Ocaña-Flaquer et al. 2010 \\
          J1521+0420 &  0.052 &  $<$9.69 &    $<$9.05 &  0.551 &         Evans et al. 2005 \\
           B21101+38 &  0.053 &  $<$9.12 &    $<$8.48 &  0.648 & Ocaña-Flaquer et al. 2010 \\
            3C390.3 &  0.057 &  $<$9.72 &    $<$9.08 & 10.225 &         Evans et al. 2005 \\
          J0119+3210 &  0.059 &     9.78 &       9.14 &  2.483 & Ocaña-Flaquer et al. 2010 \\
            4C26.42 &  0.063 &     9.51 &       8.87 &  0.925 &      Olivares et al. 2019 \\
            4C29.30 &  0.065 &     9.68 &       9.04 &  0.514 & Ocaña-Flaquer et al. 2010 \\
               OQ208 &  0.077 &    10.14 &       9.50 &  0.703 & Ocaña-Flaquer et al. 2010 \\
          B2 1707+34 &  0.082 &  $<$9.72 &    $<$9.08 &  0.474 &         Evans et al. 2005 \\
          Abell2597 &  0.083 &     9.36 &       8.72 &  1.627 &      Olivares et al. 2019 \\
              3C326 &  0.090 &     9.15 &       8.51 &  2.292 &      Nesvadba et al. 2010 \\
               3C321 &  0.095 &     9.85 &       9.21 &  3.352 & Ocaña-Flaquer et al. 2010 \\
        PKS0745-191 &  0.102 &     9.69 &       9.05 &  2.003 &      Olivares et al. 2019 \\
          PKS1559+02 &  0.105 &     9.42 &       8.78 &  7.905 & Ocaña-Flaquer et al. 2010 \\
            4C26.35 &  0.112 &  $<$9.32 &    $<$8.68 &  0.400 &     Saripalli et al. 2007 \\
             4C12.50 &  0.122 &    10.62 &       9.98 &  5.155 &    Fotopoulou et al. 2019 \\
            4C12.03 &  0.156 &  $<$9.80 &    $<$9.16 &  1.500 &     Saripalli et al. 2007 \\
        PKS2128-123 &  0.501 & $<$10.99 &   $<$10.35 &  1.895 &            Lo et al. 1999 \\
              3C285 &  0.790 &  $<$9.69 &    $<$9.05 &  1.601 &         Evans et al. 2005 \\
    B0235+164 &  0.940 & $<$11.11 &   $<$10.47 &  1.473 &            Lo et al. 1999 \\
          J0830+2410 &  0.941 & $<$11.03 &   $<$10.39 &  0.841 &            Lo et al. 1999 \\
               3C368 &  1.131 & $<$10.90 &   $<$10.26 &  1.087 &         Evans et al. 1996 \\
          J1022+3041 &  1.320 & $<$10.69 &   $<$10.05 &  0.590 &            Lo et al. 1999 \\
         MRC0114-211 &  1.410 &    11.29 &      10.65 &  3.723 &        Emonts et al. 2014 \\
              3C68.2 &  1.575 & $<$11.39 &   $<$10.75 &  0.909 &         Evans et al. 1996 \\
         MRC1017-220 &  1.768 & $<$11.06 &   $<$10.42 &  0.539 &        Emonts et al. 2014 \\
        MRC0324-228 &  1.898 & $<$11.25 &   $<$10.62 &  0.459 &        Emonts et al. 2014 \\
        MRC0152-209 &  1.921 &    11.46 &      10.82 &  0.425 &        Emonts et al. 2014 \\
         MRC0156-252 &  2.016 &    11.61 &      10.97 &  0.423 &        Emonts et al. 2014 \\
        MRC2048-272 &  2.060 &    11.58 &      10.94 &  0.485 &        Emonts et al. 2014 \\
           Spiderweb &  2.156 &    11.50 &      10.86 &  0.815 &        Emonts et al. 2014 \\
         MRC0406-244 &  2.433 & $<$11.21 &   $<$10.57 &  0.629 &        Emonts et al. 2014 \\
         MRC2104-242 &  2.491 & $<$11.06 &   $<$10.42 &  0.447 &        Emonts et al. 2014 \\

\hline
\label{tab:biblio}
\end{tabular}
\tablefoot{The values for the molecular mass listed here are corrected by our adopted CO-to-H$_2$ conversion factor. References: \citet{prandoni10}, \citet{ruffa19b}, \citet{olivares19}, \citet{ocana10}, \citet{leon02}, \citet{taniguchi90}, \citet{evans05}, \citet{saripalli07}, \citet{nesvadba10}, \citet{fotopoulou19}, \citet{lo99}, \citet{evans96} and \citet{emonts14}.}
\end{table*}

The properties of the bright sample representative of the 6C catalogue when flux limited to 3\,Jy at 151\,MHz are listed in Table~\ref{tab:bright}.

\begin{table*}[b]
\caption{CO properties of representative sample of the 6C catalogue.}
\footnotesize
\centering
\begin{tabular}{llllll}
\hline
   Galaxy  &    $z$ &  log(M$_{H_2}$) & log(L'$\rm_{CO(1-0)}$) & f$_{151 MHz}$  & reference \\
   & & (M$\rm_\odot$) & (K\kms pc$^2$) & (Jy) & \\
\hline
    NGC4696 &  0.010 &    7.95 &       7.31 & 25.618 &      Olivares et al. 2019 \\
    NGC3801 &  0.011 &    8.74 &       8.10 &  4.009 & Ocaña-Flaquer et al. 2010 \\
     IC4296 &  0.013 &    7.17 &       6.52 & 12.100 &                 This work \\
     NGC193 &  0.014 & $<$8.08 &    $<$7.44 &  5.547 &          Leon et al. 2002 \\
    NGC5490 &  0.016 & $<$8.73 &    $<$8.09 &  5.164 & Ocaña-Flaquer et al. 2010 \\
     NGC315 &  0.017 &    8.27 &       7.63 &  4.438 &                 This work \\
    NGC0383 &  0.017 &    9.23 &       8.59 & 19.224 & Ocaña-Flaquer et al. 2010 \\
 ESO443-G024 &  0.017 & $<$9.38 &    $<$8.73 &  5.801 &                 This work \\
    NGC5141 &  0.018 & $<$8.70 &    $<$8.06 &  3.044 & Ocaña-Flaquer et al. 2010 \\
 J0126-0120 &  0.018 & $<$9.55 &    $<$8.90 &  4.749 &                 This work \\
    NGC7075 &  0.018 &    7.27 &       6.62 &  3.015 &                 This work \\
     NGC741 &  0.019 & $<$8.74 &    $<$8.10 &  3.258 &         Evans et al. 2005 \\
      3C442 &  0.026 &    7.83 &       7.19 & 22.348 & Ocaña-Flaquer et al. 2010 \\
    3C129.1 &  0.027 & $<$8.48 &    $<$7.84 &  5.140 &           Lim et al. 2003 \\
      3C120 &  0.033 &    9.23 &       8.59 &  8.977 &         Evans et al. 2005 \\
     Arp187 &  0.040 &    9.60 &       8.96 & 11.539 &         Evans et al. 2005 \\
 J0758+3747 &  0.041 & $<$9.30 &    $<$8.66 & 10.341 & Ocaña-Flaquer et al. 2010 \\
      3C305 &  0.042 &    9.31 &       8.67 & 17.463 & Ocaña-Flaquer et al. 2010 \\
      3C293 &  0.045 &   10.27 &       9.63 & 17.012 &         Evans et al. 2005 \\
     NGC326 &  0.048 &    8.75 &       8.11 &  7.848 & Ocaña-Flaquer et al. 2010 \\
    4C26.42 &  0.063 &    9.51 &       8.87 &  6.033 &      Olivares et al. 2019 \\
  B21707+34 &  0.082 & $<$9.72 &    $<$9.08 &  4.610 &         Evans et al. 2005 \\
  Abell2597 &  0.083 &    9.36 &       8.72 & 15.815 &      Olivares et al. 2019 \\
      3C326 &  0.090 &    9.15 &       8.51 & 15.607 &      Nesvadba et al. 2010 \\
      3C321 &  0.095 &    9.85 &       9.21 & 21.233 & Ocaña-Flaquer et al. 2010 \\
PKS0745-191 &  0.102 &    9.69 &       9.05 & 21.554 &      Olivares et al. 2019 \\
    4C26.35 &  0.112 & $<$9.32 &    $<$8.68 &  5.001 &     Saripalli et al. 2007 \\
 J1347+1217 &  0.122 &   10.28 &       9.64 &  7.100 &         Evans et al. 2005 \\
    4C12.50 &  0.122 &   10.62 &       9.98 &  4.600 &    Fotopoulou et al. 2019 \\
\hline
\label{tab:bright}
\end{tabular}
\tablefoot{ References for M$_{\rm H_2}$ are listed in the right column: \citet{olivares19}, \citet{ocana10}, \citet{leon02}, \citet{evans05}, \citet{lim03}, \citet{nesvadba10}, \citet{saripalli07} and \citet{fotopoulou19}.}
\end{table*}

\end{appendix}

\end{document}